\documentclass[onecolumn,showpacs,aps,prd]{revtex4}
\def\ETA{\zeta}
\usepackage{graphicx}
\usepackage{dcolumn}
\usepackage{amsmath}
\usepackage{epsfig}
\RequirePackage{xspace}

\begin{document}

\def\op{{\cal O}}
\def\lsim{\mathrel{\lower4pt\hbox{$\sim$}}\hskip-12pt\raise1.6pt\hbox{$<$}\;}
\def\Dd{\psi}
\def\pp{\lambda}
\def\ket{\rangle}
\def\BAR{\bar}
\def\xba{\bar}
\def\fa{{\cal A}}
\def\fm{{\cal M}}
\def\fl{{\cal L}}
\def\ufs{\Upsilon(5S)}
\def\gsim{\mathrel{\lower4pt\hbox{$\sim$}}
\hskip-10pt\raise1.6pt\hbox{$>$}\;}
\def\ufour{\Upsilon(4S)}
\def\xcp{X_{CP}}
\def\ynotcp{Y}
\vspace*{-.5in}
\def\ETAp{\ETA^\prime}
\def\bfb{{\bf B}}
\def\fd{r_D}
\def\fb{r_B}
\def\ed{\ETA_D}
\def\eb{\ETA_B}
\def\hatA{\hat A}
\def\hatfd{{\hat r}_D}
\def\hated{{\hat\ETA}_D}
\def\D{{\bf D}}
\def\pcc{(+ charge conjugate)}

\def\uglu{\hskip 0pt plus 1fil
minus 1fil} \def\uglux{\hskip 0pt plus .75fil minus .75fil}

\def\slashed#1{\setbox200=\hbox{$ #1 $}
    \hbox{\box200 \hskip -\wd200 \hbox to \wd200 {\uglu $/$ \uglux}}}

\def\slpar{\slashed\partial}
\def\sla{\slashed a}
\def\slb{\slashed b}
\def\slc{\slashed c}
\def\sld{\slashed d}
\def\sle{\slashed e}
\def\slf{\slashed f}
\def\slg{\slashed g}
\def\slh{\slashed h}
\def\sli{\slashed i}
\def\slj{\slashed j}
\def\slk{\slashed k}
\def\sll{\slashed l}
\def\slm{\slashed m}
\def\sln{\slashed n}
\def\slo{\slashed o}
\def\slp{\slashed p}
\def\slq{\slashed q}
\def\slr{\slashed r}
\def\sls{\slashed s}
\def\slt{\slashed t}
\def\slu{\slashed u}
\def\slv{\slashed v}
\def\slw{\slashed w}
\def\slx{\slashed x}
\def\sly{\slashed y}
\def\slz{\slashed z}

\title{
%
\begin{flushright}
{AMES-HET 03-05}\\
{BNL-HET-03/23~~~}\\
\end{flushright}
%
\vskip 10mm
\large\bf
Pathways to a clean $\gamma$ ($\phi_3$): From B to Super B-factories
}

\author{David Atwood}
\affiliation{Dept. of Physics and Astronomy, Iowa State University, Ames,
IA 50011}
\author{Amarjit Soni}
\affiliation{ Theory Group, Brookhaven National Laboratory, Upton, NY
11973}

\date{\today}

\begin{abstract}

The implementation of various methods for the determination of $\gamma$
through direct CP violation arising in the interference of $b\to c$ and $b
\to u$ processes in charged as well as neutral $B$ meson decays are
considered. We show that the inclusion of $D^0$ resulting from $D^{*0} \to
D^0 + \pi^0(\gamma)$ say via $B \to K D^{*0}$ makes a significant
difference in the attainable accuracy for $\gamma$.  Both exclusive and
inclusive decays of the $B^\pm$ and $B^0$ to states containing $D^0/\bar
D^0$ followed by both inclusive and exclusive decays of the $D^0$ are
discussed. It is shown that with statistics which might be obtained at $B$
factories ($5-10 \times 10^8$ B-pairs) a $1\sigma$ determination of
$\gamma$ to $\approx \pm 5^\circ$ may be possible depending on the
efficiency of reconstruction, backgrounds and the details of the decay
amplitudes involved. The role of data from a charm factory as well as
effects of $D^0$ mixing are discussed.  Extraction of $\gamma$ with
accuracy that is roughly commensurate with the intrinsic theory error of
these methods (i.e. around 0.1\%), which is an important goal, will require
$> 10^{10}$ B-pairs, namely a Super-B Factory.

\end{abstract}


\maketitle

\section{Introduction}\label{introduction}

The B factories at KEK and SLAC have made remarkable progress in many
areas of B-physics, in particular in the extraction of Cabibbo Kobayashi
Maskawa (CKM)~\cite{ckm} parameters crucial to testing the standard model.
Indeed, the determination of $\sin 2\beta$ via $B\to J/\psi K_S$ in such a
way that there is no dependence on theoretical assumptions promises to
usher in a new era of precision tests of the CKM
paradigm\cite{belle_beta,babar_beta}.

The determination of the other two unitarity angles, $\alpha$ and
$\gamma$, without theoretical errors still presents a considerable
experimental challenge.  The corresponding CP violating effects are
somewhat harder to observe in channels sensitive to $\alpha$ and $\gamma$.
In addition, effects sensitive to $\alpha$ such as $B\to \pi\pi$ are
subject to some electroweak penguin contamination.

\begin{figure}   
\epsfxsize 2.9 in
\mbox{\epsfbox{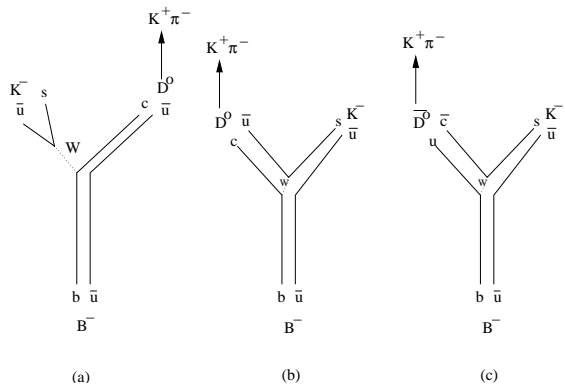}}
\caption{
Feynman diagrams for $B^-\to K^- [\D^0\to K^+\pi^-]$. Diagrams (a) and
(b) are $b\to c$ transitions where $\D^0=D^0$ while (c) is the $b\to u$
transition with $\D^0=\bar D^0$. Diagram (a) is color allowed while
(b) and (c) are color suppressed; $\bar D^0 \to K^+ \pi^-$
is Cabibbo allowed, whereas $D^0 \to K^+ \pi^-$ is
doubly Cabibbo suppressed.
}\label{feyn_1}
\end{figure}

In this paper, we will consider the application of several related methods
for determining the angle $\gamma$ through the decay $B^-\to K^-
D^0$~\cite{glw,ads,ads2,adspsi}.  The key idea of these
methods is that the decay $B^- \to K^- D^0$ and $B^- \to K^- \xba D^0$ can
interfere if the $D^0$ decays to a common hadronic final state as shown in
the generic Feynman diagram Fig.~\ref{feyn_1} where, as an example, of a
CP-non-eigenstate
(CPNES) case, the mode $B^-\to K^- [\D^0\to K^+\pi^-]$ is shown
(here we use ${\D}$ to indicate a superposition of $D^0$ and $\bar
D^0$).  
Similarly for the case of CP-eigenstate (CPES) [4] 
an example would be $D^0, \bar D^0 \to
K_S \omega$, say. These interferences involves the weak phase $\gamma$
together with unknown strong phases from the $B$ and $D$ decays. A minimum
of two modes that are common to the decays of $D^0, \bar D^0$ are needed
to solve for the weak phase $\gamma$, along with the two strong phases and
the suppressed Br($B^- \to K^- \bar D^0$) that appears to be very
difficult to measure experimentally because of serious
backgrounds~\cite{ads}.

There are a number of variations of this process which we will consider
here.  We will show that an effective method for  
the determination of
$\gamma$ is to allow the $K$ or the $D$ to assume vector or pseudo-scalar
forms. Thus the generic reaction has four variations $B^-\to K^{-}\D^0$,
$B^-\to K^{*-}\D^0$, $B^-\to K^{-}\D^{*0}$, and $B^-\to K^{*-}\D^{*0}$.
In principle, this set of reactions could be further generalized to higher
excited states of $D$ or $K$ but we will not further consider this case
here. Unless the production of such higher states is surprisingly large,
the four reactions listed are likely to be of the greatest experimental
interest.

In the cases involving $\D^{*0}$ the $\D^{*0}$ decays to a $\D^0$ with
$\pi^0$ or $\gamma$ and then the $\D^0$ decays to the hadronic final state
common to $D^0$ and $\bar D^0$.  One complication that arises in the case
of $B^-\to K^{*-}\D^{*0}$ is that there are three helicity amplitudes
which can, in principle, be separated by considering angular correlations
between the decays. Here, however, we will just consider the inclusive
data, that is without attempting to perform the angular analysis such as
that considered in~\cite{sinhas_angle}.

We will also briefly consider the
neutral $B$ decays of the same form such as
$B^0\to
K^{*0} \D^0$. 
Here we will only consider the self
tagging~\cite{desh,dunietz_selftag} cases, where the flavor of the $B$ is
determined by having a charged $K$ in the final state, for instance
$B^0\to [K^{*0}\to K^+\pi^0] \D^0$.  
For non-self-tagging modes such as
$B^0\to K_S \D^0$ one must also take into account the oscillation of the
$B^0$ in which case the main dependence of the results is on 
$\delta \equiv \beta - \alpha + \pi \equiv 2 \beta + \gamma$
as considered in~\cite{branco, kayser, sanda, deltapaper, robf}.

%
%
%
%

The key feature of these decays which allows the determination of $\gamma$
is the possibility of interference between $D^0$ and $\bar D^0$ channels.  
This can be accomplished by the selection of the common hadronic final
state of the $\D^0$ decay. The formalism given in the appendix shows how
to calculate the total rate for 
both inclusive and exclusive decays of the $D^0$.  In this case we
refer to a decay which is controlled by a single quantum amplitude as
exclusive, for instance decays to 2 pseudoscalars or a pseudoscalar and a
vector would fall into this category. Decays which are the incoherent sum
of many amplitudes are inclusive. This would include decays to two vectors
where no measurement of the polarization is made; multibody decays where
the data is integrated over all or part of phase space or states which
consist of several decay modes of different particle content. In the
formalism of the appendix, in such cases we introduce an additional
parameter $R$ which quantifies the degree of 
coherence\cite{adspsi}
associated with the
amplitudes composing the inclusive decay. This parameter, along with the
average strong phase, must also be determined from the data.

The same methodology can also be applied to inclusive decays of
the parent $B$, for instance, $B^-\to K^- \D^0 +X$ or the case $B^-\to
K^{*-}D^{*0}$ (summed over polarization). Since this approach makes no
assumption concerning the structure of the amplitude of such inclusive
decays it therefore introduces no potential model dependence into the
analysis. In contrast, the approach of~\cite{bkdpi} assumes a resonance
structure but, with this modest assumption concerning the amplitude, is
able to extract more observables since the phase space dependence is fit
in detail; 
they explicitly treat the case of $B^-\to K^-\pi^0 \D^0$,
where the $\D^0$
decays to a CP eigenstate.

The decays of the $D^0$ can be further categorized according the quark
level transition involved, it may either be Cabibbo allowed (CA), for
instance $D^0\to K^-\pi^+$; singly Cabibbo suppressed (SCS), for instance
$D^0\to K^{+*}K^-$\cite{scs} or Doubly Cabibbo suppressed (DCS) such as $D^0\to
K^+\pi^-$. Note that if $D^0\to X$ is DCS then $\bar D^0\to X$ is CA while
if $D^0\to X$ is 
SCS then so is $\bar D^0\to X$. In addition, some two body
decays such as $D^0\to \pi^+\pi^-$, $D^0\to K^+K^-$ and $D^0\to K_S\pi^0$
are CP eigenstates (CPES). The bulk of these are of the form $K_{S,L}+X$;
since they are CA and in particular about 5\% of the branching ratio is to
negative CPES (CPES-) of the form $K_S+X$~\cite{pdb}.

Section II discusses some of the B and D decays that are of interest.
In there we also spell out, in some detail, our working assumptions.
In section III we discuss the expectation for the size of the 
CP asymmetry in various modes. Section IV illustrates 
extraction of $\gamma$ for some of the combination of data
sets. In section V we re-visit the feasibility of flavor tagging
of $D^0$ versus $\bar D^0$ using their semi-leptonic decays.
Section VI deals with $D \bar D$ mixing effects.
In section
VII  we briefly discuss the use of neutral (rather than charged)
B-mesons, in conjunction with their self-tagging modes
and direct CP violation to get $\gamma$. 
Multibody B decays are briefly discussed in section VIII.
In section IX we briefly mention how these determinations
of $\gamma$ may play out at a super-B factory and we close
with a briefly summary in section X. Some details on the
formalism are relegated to the appendix.

\section{B and D Decay Modes}

In this paper, we will compare the results of various determinations
of $\gamma$ by carrying out $\chi^2$ scans assuming a certain level of input
data.

The actual number
of events observed at a given integrated luminosity will therefore depend
on phases and often even on branching ratios which are unknown. To quantify the
statistical error, we will assume that a specific fixed number of events
of each type are seen. This will have the side effect of implying a slight
variation in $\hat N_B$, (the number of $B$ mesons)$\times$(acceptance)
with the input parameters which will generally lie within a reasonable
range.  There are two advantages of adopting this approach: first of all,
the $\chi^2$ values determined for a given combination of modes will be
less dependent on the underlying unknown parameters and secondly, the
sample calculations should give an idea of the precision which can be
reached given the quota of events indicated.

In order to make contact with actual experimental conditions, we will use
a simple model for the acceptance of various modes. For each observable
particle $x$ in the final state we will assign a relative acceptance
$R_x$. In addition we will assume there is overall reduction in acceptance
of $R_{cut}$ due to the relatively hard acceptance cuts which would be
required to find the signals suggested in these methods. 
Following the discussion in\cite{bkdpi,prell}, the particular
values we will use in our 
numerical estimates are:

\begin{eqnarray}
R_{\pi^\pm}=0.95 
~~~~~~~ 
R_{K^\pm}=0.8
~~~~~~~ 
R_{\pi^0}=0.5
~~~~~~~ 
R_{\gamma}=0.5
~~~~~~~ 
R_{\eta\to 2\gamma}=0.5
\nonumber\\
R_{cut}=\frac{1}{6} - \frac{1}{3}  
\label{det_eff}
\end{eqnarray}

Perhaps the most striking initial manifestation of CP violation in this
system
will be the case where $B^-\to K^- D^0$ and the $D^0$ subsequently decays
to $K^+\pi^-$. 
As pointed out in~\cite{ads} the large decay rate $B^-\to
K^- D^0$ coupled with the Doubly Cabibbo Suppressed (DCS) decay $D^0\to
K^+\pi^-$ gives about the same amplitude as the 
color-suppressed 
decay $B^-\to K^- \bar
D^0$ coupled with the Cabibbo Allowed (CA) decay $\bar D^0\to K^+\pi^-$.
The result is that the CP violation in this combination is potentially
large. In addition, this final state is relatively easy to reconstruct so
it will likely provide the earliest evidence of CP violation from the
interference of $b\to u$ and $b\to c$ transitions.

In this paper, for simplicity,
we will assume a core data sample of 100 ${D^0\to
K^+\pi^-}$ events of this sort which are distributed equally among
four 
possible parent decays
(
$B^-\to \D^0 K^-$, 
$B^-\to \D^{*0} K^-$, 
$B^-\to \D^0 K^{*-}$ 
and
$B^-\to \D^{*0} K^{*-}$). 
In particular we will assume an initial
data sample as shown in Table~\ref{core_data_tab}. 
Note that in each
case the number of events is distributed between the 
given mode and its charge conjugate.

%
%
%
%
%
%

\begin{table}
\vspace{0.2 in}
\begin{tabular}{|c|c|c|}
\hline
Initial $B$ decay & Subsequent $D^0$ decay & Number of events \\
\hline
$B^{-}\to K^{-}D^0$ & $K^+\pi^-$ & 25 \\ 
\hline
$B^{-}\to K^{-}D^0$ & $K^{*+}\pi^-$ & 14 \\ 
\hline
$B^{-}\to K^{-}D^0$ & $K^+\pi^-+n\pi$ & 106 \\ 
\hline
$B^{-}\to K^{-}D^0$ & $CPES-$ & 827 \\ 
\hline
\end{tabular}
\caption{Initial ``core data sample" used
where the number of events is assumed to be distributed
among the given mode and its charge conjugate.
The corresponding number of events for the three other initial $B^-$
decays: 
$D^{*0}K^{-}$, 
$D^{0}K^{*-}$
and
$D^{*0}K^{*-}$
are assumed to be the same. 
}\label{core_data_tab}
\end{table}

%
%
%
%
%
%
%
%
%
%
%


In order to estimate what this implies concerning the value of $\hat N_B$,
let us assume that the interference between the two channels of $B^-\to
K^- [\D^0\to K^+\pi^-]$ is zero;  indeed this should be true on average.
In Table~\ref{branching_ratio_A} discussed below, we list the branching
ratios which we use for various $B$ and $D$ decays. Using these numbers,
the sum of the two channels neglecting interference is $26\times 10^{-8}$.
Since only charged B's are being used here, for the needed 25 K $\pi$
events, it will require $N_\Upsilon = (3 - 6) \times 10^8$, where we have
included $R_{cut} = [\frac{1}{6} - \frac{1}{3}]$; corresponding to 300-600
$fb^{-1}$.  To estimate the number of events for the other modes we will
combine the above acceptances for each of the final state particles.


We will assume that the product of these two quantities is roughly the
same between the different combinations of $K^*$ and $D^{*0}$ in order to
give the same number of events in each case, the other modes tend to have
larger branching ratios but will likely have smaller acceptances.


The $K^+ \pi^-$ mode alone given 
in Table~\ref{core_data_tab} is not sufficient to determine 
$\gamma$, in general at least one additional 
decay mode of $D^0$
is required. Another 
prominent DCS mode which could be used is $K^{*+}\pi^-$\cite{ads,ads2}. 
In principle, this
is a 3 body mode and can interfere to some extent with other quasi 2 body
modes; however, the $K^*$ is sufficiently narrow that taking it as a two
body mode should be a good approximation.

In the CA channel, this mode is about 1.5 times that of $D^0\to K^-\pi^+$ and
we will assume that this is roughly true for the overall $B^-\to
K^-[\D^0\to K^{*+} \pi^-]$.  The $K^{*+}$ has three dominant decay modes,
each of 
which has branching ratio $\sim\frac13$:  $K_L\pi^+$, $K_S\pi^+$ and
$K^+\pi^0$.  In the case of $K_L\pi^+$ the $K_L$ will be hard to detect
and we will assume that the acceptance is negligible.  For $K_S\pi^+$ the
$K_S$ decays $\frac23$ to $\pi^+\pi^-$ giving a final state which is
likely to have high acceptance. There is, however a background from the CA
decay $D^0\to \pi^+[K^{*-}\to K_S\pi^-]$ so that some cuts may be needed.
We will assume that this channel has an acceptance reduced by $\frac12$
due to the cuts giving a total of $\frac13$ taking into account
$Br(K_S\to\pi^+\pi^-)$. There is no such problem with $K^{*+}\to K^+\pi^0$
but the neutral pion in the final state will reduce the acceptance by
$0.5$.  

Overall then, the acceptance for $B^-\to K^-[\D^0\to K^{*+}\pi^-]$ is
0.22. The total number of events then should be .55 times the case of
$\D^0\to K^+\pi^-$ so we will assume 14 events for each $D^{(*)0}K^{(*)-}$
combination as shown in Table~\ref{core_data_tab}.


Another class of modes are CP eigenstates (CPES) as considered
in~\cite{glw} which we will refer to as GLW modes. Such modes, excluding
those with $K_L$ have a branching ratio of about 5\%.

In Table~\ref{cpes_tab} we list some such modes with the central value of
the branching ratio from~\cite{pdb} and acceptance for $B^-\to K^-[D^0\to
CPES]$ estimated as above. The product of these two gives an effective
branching ratio; for CP=-1 (CPES-) states it is 1.6\% while for CP=+1
(CPES+) this is .32\%. In our estimates we will consider the CPES- states
only in which case if there 
are 25 $K^+\pi^-$ events there should be about 827 
CPES- events as indicated in Table~\ref{core_data_tab}.

\begin{table}
\vspace{0.2 in}
\begin{tabular}{|c|c|c|c|}
\hline
CP Eigenstate  & Br [\%]
& Est. Acceptance
 & eff. Br [\%]
\\ \hline
\hline

$K_S\pi^0$           &  1.14            & 0.360             & 0.410

\\ \hline

$K_S 
[\eta\to \pi^+\pi^-\pi^0; \pi^+\pi^-\gamma]$  
                              &  0.38  & 0.231              & 0.088
\\ \hline

$K_S \rho^0$          &  0.73             & 0.648            & 0.473

\\ \hline

$K_S [\omega\to 3\pi]$           &  1.1     & 0.324 &  0.356

\\ \hline

$K_S 
[\eta^\prime\to \pi^+\pi^-\gamma; \pi^+\pi^-\eta]$ 
                           &  0.93     & 0.188            &  0.175
\\ \hline

$K_S [\phi\to K^+K^-]$             &  0.47 & 0.261           & 0.123

\\ \hline

Total CPES-          &  4.75             &                & 1.625

\\ \hline
\hline

$K_S [f_2(1270)\to \pi^+\pi^-]$   & 0.22  & 0.549           & 0.121

\\ \hline

$\pi^+\pi^-$           & 0.14  &                0.648         & 0.091
\\ \hline

$K^+K^-$              & 0.412 &                0.512        & 0.211

\\ \hline

$K_S K_S $            & 0.03 &                 0.567         & 0.017

\\ \hline

Total CPES+          & 0.58 &     0.55                     & 0.319

\\ \hline

\end{tabular}
\caption{
This table consists of a list of some of the CPES decays of the 
$D^0$. For each mode, the branching ratio (Br) is given as the central
value in~\cite{pdb} and the acceptance estimated as discussed in the
text. The effective branching  ratio is the product of these two.
}\label{cpes_tab}
\end{table}

%
%


%
%
%
%

Singly Cabibbo suppressed modes which are not CP eigenstates such as
$D^0\to K^{*+}K^-$ have been considered in~\cite{scs}.
These have the
property that the decay of $D^0$ to the charge conjugate   
i.e. 
$D^0\to K^{+}K^{*-}$ will also be possible.
We will therefore be able to observe four different reactions:

\begin{eqnarray}
B^- &\to& K^- [\D^0\to K^{*+} K^-] \nonumber\\
B^+ &\to& K^+ [\D^0\to K^{*-} K^+] \nonumber\\
B^- &\to& K^- [\D^0\to K^{*-} K^+] \nonumber\\
B^+ &\to& K^+ [\D^0\to K^{*+} K^-]
\label{fourscs} 
\end{eqnarray}

\noindent Note that the strong phase difference in the first two decays is
the opposite of that in the second two. In fact, both of these modes taken
together will allow the determination of $\gamma$.  The central values of
the two $D$ branching ratios~\cite{pdb} are 0.38\% for $D^0\to K^{*+}K^-$
and 0.22\% for $D^0\to K^{*-}K^+$.  Estimating the number of
such events in our data sample as above we find about 50 in each case.

%
%
%
%


Another data set which might be obtained is by using semi-leptonic (SL)
decays of the $D^0$ to directly monitor the $b\to u$ transition $B^-\to
K^{(*)-} \bar D^{(*)0}$. This is likely to be difficult as discussed
in~\cite{ads} as it requires flavor tagging of $D^0$ versus $\bar D^0$
which has non-trivial backgrounds from semi-leptonic decays of the $B$.
In Section V, we will discuss some possible strategies for
overcoming this potential difficulty. To illustrate possible usefulness of
this determination, we will then show the effect on $\gamma$
extraction assuming that data of this sort gives a 1-$\sigma$ error on
$Br(B^-\to K^- D^0)$ ranging from $200\%$ to $10\%$. Thus the
implementation of the original GLW\cite{glw} method to obtain $\gamma$,
may be envisioned with the use of CPES along with flavor tagging through
the use of SL data.

Additional information to reconstruct $\gamma$ may also be obtained
through the use of a $\psi(3770)$ charm factory to measure the strong
phases in $D^0$ decays~\cite{adspsi,sofferun,ss1,ggr}. As discussed
in~\cite{adspsi} this can provide additional information 
so that $\gamma$ can also be determined through inclusive modes.
Even if charm factory data is not available, if we combine inclusive 
modes with other exclusive modes, there is enough information 
to determine
$\gamma$ and we shall explore a number of such possibilities below.

In particular, as an example of such a mode we will consider the
combination of modes $K^+\pi^-+ n\pi$. 
Most likely each of these modes will be detected separately so additional
information~\cite{ads2,scs2} may  be obtained 
by separating such events according to particle content or by analyzing
the phase space distributions. In the early stages, however, such
analysis may not give significant gains over treating these modes as 
inclusive ones.

In Table~\ref{knp_tab} we show the acceptance and effective branching
ratios for various modes of this form. Note that we subtract
$\pi^+[K^{*-}\to K^-\pi^0]$ from the sample to avoid double counting. 
Again
we will assume that the DCS modes are related by a factor of
$\sin^4\theta_C$. 
Because this is an
inclusive state, we also need to specify the coherence factor $R_F$
defined in the appendix. We will assume $R_F=0.51$
for our numerical illustration
which is the value
determined in the model discussed in~\cite{adspsi}.
The branching ratio to these modes
via the CA channel are about 7 times that of $\bar D^0\to K^+\pi^-$ with
a
similar ratio being true for the DCS analogs. Assuming the acceptance
is
roughly 1/2 of the two body decay, the number of events in this sample
(106) is
thus as shown in Table~\ref{core_data_tab}.

\begin{table}
\vspace{0.2 in}
\begin{tabular}{|c|c|c|c|}
\hline
CP Eigenstate  & Br [\%] (CA)
& Est. Acceptance
 & eff. Br [\%]
\\ \hline
\hline
$K^-\pi^0\pi^+$             & 11.1  &  .304  &  3.37  \\ \hline
$K^-\pi^+\pi^+\pi^-$        & 7.5   &  .549  &  4.12 \\ \hline
$K^-\pi^+\pi^+\pi^-\pi^0$   & 4.0  &   .275  &  1.10 \\ \hline
Total                       & 22.6 &    &       8.59 \\ \hline
\end{tabular}
\caption{
The Cabibbo allowed (CA) branching ratio 
for various decays of the form $D^0\to
K^-\pi^++n\pi$ are shown together with the acceptance as discussed above
and the effective branching ratio. We will assume that the corresponding
doubly-Cabibbo-suppressed (DCS) modes are equal to these 
multiplied by a factor of
$\sin^4\theta_C$. In the case of $K^-\pi^0\pi^+$ we have subtracted the
portion which is produced via the $K^{*-}\pi^+$ channel to avoid double
counting with that mode. 
}\label{knp_tab}
\end{table}


In summary, we will investigate the determination of $\gamma$ through
considering various combinations of the following types of data
shown in Table~\ref{core_data_tab}:

\begin{enumerate}

\item\label{core} $\D^0\to K^+\pi^-$. 

\item\label{kstar} 
$\D^0\to K^{*+}\pi^-$. 

\item\label{glw} 
GLW modes, i.e. $\D^0\to CPES$.

\item\label{inclusive}
Inclusive 
decays of the form
$D^0\to K^+\pi^-+n\pi$.

\item\label{kkstar}
Singly Cabibbo Suppressed $\D$ decays such as 
$D^0\to K^{*\pm}K^{\mp}$.

\item\label{semilep}
A determination of $B^-\to K^- \bar D^0$ via a semi-leptonic tag of
$\bar D^0$.

\item\label{charmfactory}
A determination of strong phases and the coherence factor,
$R_F$, from a $\psi(3770)$ charm factory

\end{enumerate}

These will be combined with the following decays of the initial $B$ meson:

\begin{enumerate}
\item
$B^-\to K^{(*)-}\D^{(*)0}$
\item
$B^-\to K^- \D^0 +X$
\item
$B^-\to  [K^{*-}\to K^-\pi^0] \D^0 +X$
\end{enumerate}

The accuracy in determining $\gamma$ will, in general, depend on the value
of $\gamma$, the strong phase of the $B$ decay and the strong phase of
the $D$ decays. Clearly, these quantities are unknown and cannot be
calculated.  We are thus driven to consider sample calculations with
arbitrarily chosen strong phases in order to project how the measurement
will proceed.

%
%
%
%
%
%
%
%
%

For our sample calculations we will assume that $\gamma=60^\circ$ which is
consistent with current data~\cite{ckm_fit}.  The strong phase differences
for the various $B$ and $D$ decays we will 
use in our illustrative calculations are chosen
completely arbitrarily and are given in
Table~\ref{strong_phase_A}.  The table gives the strong phase difference
between the B decay involving a $D^0$ in the first column and the
corresponding decay involving $\bar D^0$ in the second column.  Note that
in the case of $B^-\to K^{*-} \D^{*0}$ there are three different helicity
amplitudes, helicity $h=+1$, $0$, and $-1$, hence there are three
different strong phases.

\begin{table}
\vspace{0.2 in}
\begin{tabular}{|c|c|c|}
\hline
Decay 1& Decay 2  &  Phase Difference   \\
\hline

\hline
$B^-\to D^0 K^-$     & $B^-\to \bar D^0 K^-$    &  $-50^\circ $ \\ \hline
$B^-\to D^{*0} K^-$  & $B^-\to \bar D^{0*} K^-$ &  $-10^\circ $ \\ \hline
$B^-\to D^0 K^{*-}$  & $B^-\to \bar D^0 K^{*-}$ &  $+30^\circ $ \\ \hline
$B^-\to D^{*0} K^{*-}_{h=+1}$& $B^-\to\bar D^{*0}K^{*-}_{h=+1}$&
$+70^\circ $ \\
$B^-\to D^{*0} K^{*-}_{h=0}$& $B^-\to\bar D^{*0}K^{*-}_{h=+1}$&
$+110^\circ$ 
\\
$B^-\to D^{*0} K^{*-}_{h=-1}$& $B^-\to\bar D^{*0}K^{*-}_{h=+1}$&
$+150^\circ
$ \\
\hline
\hline
$B^0\to D^0 K^{*0}$  & $B^-\to \bar D^0 K^{*0}$ &  $130^\circ $ \\ \hline
$B^0\to D^{*0} K^{*0}_{h=+1}$& $B^-\to\bar D^{*0}K^{*0}_{h=+1}$&$+50^\circ$\\
$B^0\to D^{*0} K^{*0}_{h=0}$& $B^-\to\bar D^{*0}K^{*0}_{h=+1}$&$+10^\circ$\\
$B^0\to D^{*0} K^{*0}_{h=-1}$& $B^-\to\bar D^{*0}K^{*0}_{h=+1}$&$-30^\circ
$ \\
\hline
\hline
$B^-\to D^0 K^- + X$  & $B^-\to \bar D^0 K^- + X$ & $45^\circ$  $R=0.7$
\\ \hline
$B^0\to D^0 K^- + X$  & $B^0\to \bar D^0 K^- + X$ & $135^\circ$ $R=0.7$ 
\\ 
\hline
\hline
$D^0\to K^+\pi^-$ & $\bar D^0\to K^+\pi^-$ &  $ +120^\circ $ \\ \hline
$D^0\to K^+\pi^-+n\pi$ & $\bar D^0\to K^+\pi^-+n\pi$ &  $+90^\circ 
$($R=0.5$) \\ 
\hline
$D^0\to K^{*+}\pi^-$ & $\bar D^0\to K^{*+}\pi^-$ &  $+60^\circ $ \\ \hline
$D^0\to K^{*+}K^-$ & $\bar D^0\to K^{*+}K^-$ &  $ +30^\circ $ \\ \hline
\end{tabular}
\caption{    
The strong phase differences used in our sample calculations. The strong
phase difference is between decay 1 involving $D^0$ and decay 2 involving
$\bar D^0$.
}\label{strong_phase_A}
\end{table}

%
%
%
%
%
%

In order to carry forward our analysis we will need to make some
assumptions concerning branching ratios of decays which 
have not been 
measured yet.  In Table~\ref{branching_ratio_A}  we summarize the branching
ratios for the decays which we use. The decays which are measured are
taken from~\cite{pdb} and indicated by $(*)$ and the ones which we
estimate are indicated by $(\dag)$.

\begin{table}
\vspace{0.2 in}
\begin{tabular}{|c|c|c|}
\hline
Decay   & Branching Ratio & Status  \\
\hline
\hline
$\bar D^0\to K^+\pi^- $ &  $3.80\%            $      &$*$ \\ \hline
$     D^0\to K^+\pi^- $ &  $1.48\times 10^{-4}$      &$*$ \\ \hline
$\bar D^0\to K^{*+}\pi^- $ &  $6.0\%          $      &$*$ \\ \hline
$     D^0\to K^{*+}\pi^- $ &  $1.7\times 10^{-4}$    &$\dag$ \\ \hline
$\bar D^0\to K^+\pi^-+n\pi $ &  $25\%$               &$\dag$ \\ \hline
$     D^0\to K^+\pi^-+n\pi $ &  $7\times 10^{-4}$    &$\dag$ \\ \hline
$     D^0\to CPES- $         &  $5\%$                &$\dag$ \\ \hline
$\bar D^0\to K^{*-}K^- $ &  $0.380\%          $      &$*$ \\ \hline
$     D^0\to K^{*-}K^- $ &  $0.220\%          $      &$*$ \\ \hline
$B^-\to K^- D^0$             &  $3.7\times 10^{-4}$  &$*$ \\ \hline
$B^-\to K^{*-} D^0$             & $6.1\times 10^{-4}$  &$*$ \\ \hline
$B^-\to K^{-} D^{*0}$           & $3.6\times 10^{-4}$  &$*$ \\ \hline
$B^-\to K^{*-} D^{*0}$          & $7.2\times 10^{-4}$  &$*$ \\ \hline
$B^-\to K^- \bar D^0$             & $5.4\times 10^{-6}$  &$\dag$\\ \hline
$B^-\to K^{*-} \bar D^0$          & $9.0\times 10^{-6}$  &$\dag$ \\ \hline
$B^-\to K^{-} \bar D^{*0}$        & $5.3\times 10^{-6}$  &$\dag$ \\ \hline
$B^-\to K^{*-} \bar D^{*0}$       & $1.06\times 10^{-5}$  &$\dag$ \\ \hline
$\bar B^0\to [\bar K^{*0}\to K^-\pi^+] D^0$    & $4.6\times 10^{-5}$
&$\dag$ \\ \hline
$\bar B^0\to [\bar K^{*0} \to K^-\pi^+]
D^{*0}$ & $5.5\times 10^{-5}$  &$\dag$ \\ \hline
$\bar B^0\to [\bar K^{*0}  \to K^-\pi^+]
\bar D^0$  & $6.0\times 10^{-6}$  &$\dag$ \\ \hline
$\bar B^0\to [\bar K^{*0}   \to K^-\pi^+]
\bar D^{*0}$  & $7.0\times 10^{-6}$  &$\dag$ \\ \hline
\end{tabular}
\caption{Branching ratios for various 
$B$ and $D$ decays. Those from~\cite{pdb} are indicated by $*$ while those
which have been estimated have been indicated by $\dag$.
}\label{branching_ratio_A}
\end{table}

\section{CP Asymmetries}\label{cp_asymmetries}

For a given decay of the form $B^-\to K^{(*)-}[\D^{(*)0}\to F]$ where 
$F$ is in general an exclusive or inclusive final state, let us denote

\begin{eqnarray}
d&=&Br(B^- \to K^{(*)-}[D^{(*)0}\to F] \nonumber\\
\bar d
&=&Br(B^+ \to K^{(*)+}[D^{(*)0}\to \bar F] \nonumber\\
\end{eqnarray}

\noindent where in the case of $\D^{*0}$ it cascades down to $\D^0$ before a
$\D^0\to F$ decay.  
In terms of the $B$ and $D$ branching ratios and the strong and weak
phases, these quantities are given by

\begin{eqnarray}
d
&=& 
ac_F + 
b\bar c_F  
+2R_F\sqrt{ac_Fb\bar c_F}
\cos(\zeta_B+\zeta_F+\gamma)
\nonumber\\ 
\bar d
&=& 
ac_F + 
b\bar c_F  
+2R_F\sqrt{ac_Fb\bar c_F}
\cos(\zeta_B+\zeta_F-\gamma)
\label{dsystem}
\end{eqnarray}

\noindent
where 
$a=Br(B^-\to K^{(*)-}D^{(*)0})$,
$b=Br(B^-\to K^{(*)-}\bar D^{(*)0})$,
$\zeta_B$ 
is the strong phase difference of the $B^-$ decay,
$\zeta_{F}$ is the strong phase difference for the $D^0$ decay 
and $R_F$ is the coherence factor as defined in~\cite{adspsi};
in particular, for exclusive states $R_F = 1$.
In the appendix, these expressions are generalized to the case where 
$D^0\bar D^0$ mixing is present.

For each decay, let us define 

\begin{eqnarray}
d_{av} &=& (d+\bar  d)/2 \nonumber\\
A_{CP} &=& (d-\bar  d)/(d+\bar  d)
\end{eqnarray}

\noindent 
thus, $A_{CP}$ is the CP violating asymmetry.


\begin{table}
\vspace{0.2 in}
\begin{tabular}{|c|c|c|c|c|c|c|}
\hline
\ &   $K^+\pi^-$ & $K^{*+}\pi^-$ & $K^+\pi^-+n\pi$ & $CPES-$ &
$K^{*+} K^-$ & $K^{*-} K^+$
\\
\hline
\hline
$B^-\to K^- D^0$
& 29 & 53 & 165 & 1646 
& 70  & 51  
\\ \hline
$B^-\to K^{*-} D^0$
& 28 & 64 & 219 & 2636  
& 74  &  44
\\ \hline
$B^-\to K^- D^{*0}$ 
&22 & 47 & 147 & 1530   
& 66  &  53
\\ \hline
$B^-\to K^{*-} D^{*0}$  
&40 & 53 & 244 & 3590 
& 44  & 45
\\ \hline
\end{tabular}
\caption{The calculated branching ratio, 
$d_{av}=(d+\bar d)/2$ in units of $10^{-8}$
for the combination of the various
$B^-\to
K^{(*)-}D^{(*)0}$ decays in the left column with the $D^0$ decays in the
top row using the parameters from Table~\ref{strong_phase_A},
where the strong phases were chosen arbitrarily. 
As indicated,
the results are averaged over each decay
chain and its charge conjugate.
}\label{d_av_tab}
\end{table}


\begin{table}
\vspace{0.2 in}
\begin{tabular}{|c|c|c|c|c|c|c|}
\hline
\ &   $K^+\pi^-$ & $K^{*+}\pi^-$ & $K^+\pi^-+n\pi$ & $CPES-$ &
$K^{*+} K^-$ & $K^{*-} K^+$
\\
\hline
\hline
$B^-\to K^- D^0$
& 22 & 12 & 63 & 905 & 13  & 9  
\\ \hline
$B^-\to K^{*-} D^0$
& 21 & 14 & 83 & 1450  & 13  &  8
\\ \hline
$B^-\to K^- D^{*0}$ 
&17 & 11 & 56 & 842   & 12  &  10
\\ \hline
$B^-\to K^{*-} D^{*0}$  
&30 & 12 & 93 & 1974 & 8  & 8
\\ \hline
\end{tabular}
\caption{
The branching ratio $d_{av}$, in units of $10^{-8}$,
as in Table~\ref{d_av_tab} with the
acceptance (not
including $R_{cut}$) folded in.
}\label{d_av_tab_accept}
\end{table}


\begin{table}
\vspace{0.2 in}
\begin{tabular}{|c|c|c|c|c|c|c|}
\hline
\ &   $K^+\pi^-$ & $K^{*+}\pi^-$ & $K^+\pi^-+n\pi$ & $CPES-$ &
$K^{*+} K^-$ & $K^{*-} K^+$ 
\\
\hline
\hline
$B^-\to K^- D^0$
& -58.2\% & -8.1\% & -17.9\% & -17.1\% & -17.1\%   & 14.3\%
\\ \hline
$B^-\to K^{*-} D^0$
& -54.3\% & -63.3\% & -30.1\% & +11.6\% & -15.0\%  & 13.4\%
\\ \hline
$B^-\to K^- D^{*0}$
&-76.9\% & -39.4\% & -30.2\% & -4.1\% &  -12.1\%   & 8.1\%
\\ \hline
$B^-\to K^{*-} D^{*0}$
&+58.4\% & -13.5\% & 10.7\% & +15.8\% &  +14.6\%   & -7.2\%
\\ \hline
\end{tabular}
\caption{
The CP asymmetry $A_{CP}=(d-\bar d)/(d+\bar d)$ corresponding to the
results in Table~\ref{d_av_tab} is given.
}\label{asy_tab}
\end{table}


\begin{table}
\vspace{0.2 in}
\begin{tabular}{|c|c|c|c|c|c|c|}
\hline
\ &   $K^+\pi^-$ & $K^{*+}\pi^-$ & $K^+\pi^-+n\pi$ & $CPES-$ &
$K^{*+} K^-$ & $K^{*-} K^+$ 
\\
\hline
\hline
$B^-\to K^- D^0$
& 0.92 & 25.88 & 1.70 & 0.19 & 4.40   & 8.63
\\ \hline
$B^-\to K^{*-} D^0$
& 1.09 & 0.35 & 0.45 & 0.25 & 5.41  & 11.39
\\ \hline
$B^-\to K^- D^{*0}$
& 0.69 & 1.23 & 0.67 & 3.50 &  9.31   & 25.88
\\ \hline
$B^-\to K^{*-} D^{*0}$
& 0.66 & 9.32 & 3.22 & 0.10 &  9.60   &  38.58
\\ \hline
\end{tabular}
\caption{
The number $\hat N=9/(d_{av}A_{CP}^2)$ of $B \bar B$ pairs 
needed to see a $3-\sigma$ signal 
of CP-violation
in units of $10^8$; detection efficiencies, acceptances etc
are {\it not} included in these numbers but are included in
Table~\ref{n_tab}.  
}\label{n_tilde_tab}
\end{table}



\begin{table}
\vspace{0.2 in}
\begin{tabular}{|c|c|c|c|c|c|c|}
\hline
\ &   $K^+\pi^-$ & $K^{*+}\pi^-$ & $K^+\pi^-+n\pi$ & $CPES-$ &
$K^{*+} K^-$ & $K^{*-} K^+$
\\
\hline
\hline
$B^-\to K^- D^0$
& 6.05 & 508 & 22.4 & 1.73 & 122   & 240
\\ \hline
$B^-\to K^{*-} D^0$
& 7.17 & 7.95 & 5.92 & 2.27 & 150  & 306
\\ \hline
$B^-\to K^- D^{*0}$
& 4.53 & 28.0 & 8.82 & 31.8 &  259   & 722
\\ \hline
$B^-\to K^{*-} D^{*0}$
& 4.34 & 211 & 42.4 & 0.9 &  267  &  1083
\\ \hline
\end{tabular}
\caption{
The number $N=9/(d_{av}A_{CP}^2)$ of $B \bar B$ pairs
needed to see a $3-\sigma$ signal
of CP-violation
in units of $10^8$ taking into account the detection efficiencies,
acceptances discussed in the
text and 
assuming $R_{cut}=0.2$; see Eq.~\ref{det_eff}. 
}\label{n_tab}
\end{table}
                                                                                

\begin{table}
\vspace{0.2 in}
\begin{tabular}{|c|c|c|c|c|}
\hline
\ &   $K^+\pi^-$ & $K^{*+}\pi^-$ & $K^+\pi^-+n\pi$ & $CPES-$ \\
\hline
\hline
$B^-\to K^- D^0$
& 26 & 38 & 161 & 1877
\\ \hline
$B^-\to K^{*-} D^0$
& 43 & 64 & 268 & 3095  
\\ \hline
$B^-\to K^- D^{*0}$
&25 & 38 & 158 & 1827
\\ \hline
$B^-\to K^{*-} D^{*0}$
&51 & 76 & 315 & 3653 
\\ \hline
\end{tabular}
\caption{The value of $d_{av}$ is given using $\gamma=60^\circ$ and the 
branching ratios in Table~\ref{branching_ratio_A} but with all of the
strong phases taken at random.
}\label{d_av_tabB}
\end{table}

\begin{table}
\vspace{0.2 in}
\begin{tabular}{|c|c|c|c|c|}
\hline
\ &   $K^+\pi^-$ & $K^{*+}\pi^-$ & $K^+\pi^-+n\pi$ & $CPES-$ \\
\hline
\hline
$B^-\to K^- D^0$
& 53.4\% & 47.7\% & 47.5\% & 14.6\%
\\ \hline
$B^-\to K^{*-} D^0$
& 53.4\% & 47.7\% & 47.5\% & 14.6\%
\\ \hline
$B^-\to K^- D^{*0}$
& 53.4\% & 47.7\% & 47.5\% & 14.6\%
\\ \hline
$B^-\to K^{*-} D^{*0}$
&36.8\% & 33.0\% & 23.0\% & 10.4\%
\\ \hline
\end{tabular}
\caption{
The r.m.s. average of $|A_{CP}|$ corresponding to the results in
Table~\ref{d_av_tabB}.
}\label{asy_tabB}
\end{table}

\begin{table}
\vspace{0.2 in}
\begin{tabular}{|c|c|c|c|c|}
\hline
\ &   $K^+\pi^-$ & $K^{*+}\pi^-$ & $K^+\pi^-+n\pi$ & $CPES-$ \\
\hline
\hline
$B^-\to K^- D^0$
& 44.3\% & 39.3\% & 39.1\% & 12.0\%
\\ \hline
$B^-\to K^{*-} D^0$
& 44.3\% & 39.3\% & 39.1\% & 12.0\%
\\ \hline
$B^-\to K^- D^{*0}$
& 44.3\% & 39.3\% & 39.1\% & 12.0\%
\\ \hline
$B^-\to K^{*-} D^{*0}$
&30.3\% & 27.1\% & 18.8\% & 8.5\%
\\ \hline
\end{tabular}
\caption{
The results as in Table~\ref{asy_tabB} averaged over all values of
$\gamma$ (from 0 to $2\pi$).
}\label{asy_tabBx}
\end{table}

In Table~\ref{d_av_tab} we show the values of $d_{av}$ which result from
the parameters of our sample calculation
in units of $10^{-8}$. 
In
Table~\ref{asy_tab} we show the corresponding CP asymmetry $A_{CP}$. 
In an {\it ideal} detector, the number of $B\bar B$ pairs required to
observe
the signal of CP violation (for definiteness, with a
3 $\sigma$ statistical error) is:

\begin{eqnarray}
\hat N &=& \frac{9}{d_{av}A_{CP}^2}
\end{eqnarray}

\noindent Tables~(\ref{n_tilde_tab},\ref{n_tab}) 
indicate the number of $B \bar B$
pairs which would be required to observe a $3-\sigma$ signal of CP
violation in these modes; Table~\ref{n_tab} includes efficiencies
and acceptance factors as given in Eq.~\ref{det_eff}.
From these tables it can be seen that with a $(1-10)\times
10^8$ $B$'s CP violation can be seen in several individual channels. In
Table~\ref{d_av_tabB} we show the values of $d_{av}$ where we average
over strong phases but keep $\gamma=60^\circ$; likewise in
Table~\ref{asy_tabB} we show the r.m.s averaged values of $|A_{CP}|$.
Comparing these to the specific results in 
Table~\ref{asy_tab}, obtained by assuming arbitrarily assigned
values for strong phases, we see that the
two approaches tend to give similar results.
In
Table~\ref{asy_tabBx} we show the r.m.s asymmetries where we have also
averaged over $\gamma$ where $\gamma$ ranges from $0$ to $2\pi$.

\section{Extracting $\gamma$ \label{gamma}}

We will now consider various strategies to determine $\gamma$ assuming we
experimentally observe the results given in
Tables~(\ref{d_av_tab},\ref{asy_tab}) with the number of reconstructed
events as discussed above.

First, let us consider in isolation the case of $B^-\to K^- [D^0\to
K^+\pi^-]$. This rate, together with its charge conjugate gives us two
distinct observables which are determined in terms of four unknown
parameters: $\zeta_{KD}$, $\zeta_{K^+\pi^-}$, $b(KD)$ and $\gamma$. The
two strong phases enter as the sum
$\zeta_{tot}=\zeta_{KD}+\zeta_{K^+\pi^-}$ so in effect there are only
three parameters $\{\zeta_{tot}$, $b(KD)$, $\gamma\}$. Still we cannot
expect to reconstruct $\gamma$ but, as discussed in~\cite{ads2}, this data
gives a  bound on $\sin^2\gamma$.

\begin{figure}   
\epsfxsize 2.9 in
\mbox{\epsfbox{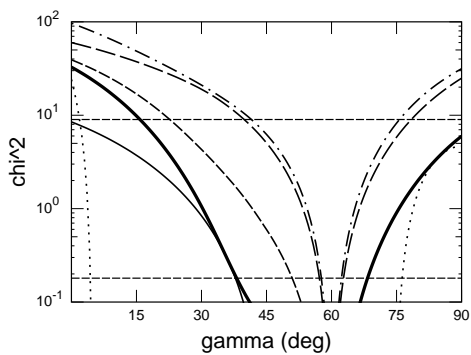}}
\caption{
The minimum value of $\chi^2$ is shown as a function of 
$\gamma$ for various combinations of data in the 
sample calculation. 
The thin solid line shows the result  using just 
$B^-\to K^-[\D^0\to K^+\pi^-]$  data.
The dotted line shows the result using just the 
$B^-\to K^-[\D^0\to CPES-]$  data.
The thick solid curve shows the result taking both 
$B^-\to K^-[\D^0\to K^+\pi^-]$   and
$B^-\to K^-[\D^0\to CPES-]$  data together.
The dashed line shows the result using
$B^-\to K^{(*)-}[\D^{(*)0}\to K^+\pi^-]$.
In the dash dotted curve, all four of the initial $B^-$ decays 
where the $\D$ decays to the same two final states are considered. Thus 
the dashed dotted curve results from taking together data of the form 
$B^-\to K^{(*)-}[\D^{(*)0}\to K^+\pi^-]$   and
$B^-\to K^{(*)-}[\D^{(*)0}\to CPES-]$.
The long dashed curve only includes data from two parent $B^-$ decays, 
i.e. $B^-\to K^{-} D^0$ as well as $B^- \to K^-
[\D^{*0} \to \D^0 + \pi^0(\gamma)]$ with either
of the two $\D^0$ decaying to $K^+ \pi^-$ as well as $CPES-$.  
The upper horizontal dashed line indicates the 3-sigma level
determination of $\gamma$ with 
the luminosity required to give the results in
Table~\ref{core_data_tab} which corresponds to current B-factories while
the lower horizontal dashed line corresponds to a luminosity 50 times
greater which may be achieved at future high luminosity B factories.
}\label{chisq_1}
\end{figure}

To illustrate this, in Fig.~(\ref{chisq_1}) the thin solid line shows the
minimum value of $\chi^2$ as a function of $\gamma$. For each value of
$\gamma$ we minimize with respect to the other parameters $\{\zeta_{tot}$,
$b(KD)\}$.  In the region of the graph where $\gamma>45.6^\circ$, $\chi^2$
vanishes indicating that this is the lower bound within the first
quadrant. Note that this and all similar graphs we will discuss are
periodic with respect to $\gamma\to\pi\pm\gamma$ so, for instance, in the
second quadrant the bound is $\gamma<180^\circ - 45.6^\circ$ etc..

Clearly, in this case, the data is too meager to provide a useful
bound. The $3\sigma$ bound (i.e. where $\chi^2\approx 9$) is  
only slightly above 0.

We can also consider the bound on $\gamma$ obtained from the decay $D^0\to
CPES-$ in isolation.  There are more events of this type but the power
this data to bound $\gamma$ is not much greater since $A_{CP}$ is
smaller (in general we expect the analyzing power of a particular mode to
be $\sim A_{CP}^2$). The minimum $\chi^2$ in this case is shown with the
dotted curve.
Notice that taken in isolation it seems to be worse than
even the single $D^0 \to K^+ \pi^-$ (CPNES) mode.

Of course both of these two data sets depend on the common parameter $b$
and if we have both sets of data together we obtain the results shown with
thick solid curve which is an improvement on each of the data sets taken
in isolation; 
in fact this thick solid curve gives a 3 $\sigma$
bound of $\gamma > 16^\circ$.  
As discussed in~\cite{ads} since there the number of
equations and observables is the same, 
there are ambiguous solutions which leads to the
$\chi^2$ value being small over an extended range.

To improve the situation, we can also use data from all four decays of the
form $B^-\to K^{(*)-} D^{(*)0}$.  Note that each of these modes will have
a different unknown value of $b$ and $\zeta$. In addition, the decay mode
$K^{*-}D^{*0}$ has three polarization amplitudes which we will take
into account by introducing a coherence factor $R_F$ into the fit since we
are assuming that we are only observing the sum and we do not consider the
additional information that could be determined from the angular
distributions of the decays of the vectors as discussed
in~\cite{sinhas_angle}.  
If we consider the single decay $D^0\to K^+\pi^-$ we obtain the results
shown by the dashed line which in this case gives a $3-\sigma$ bound on
$\gamma$ of $\gamma>23^\circ$. The dot dash curve shows the result where
we have both the $D^0\to K^+\pi^-$ and $D^0\to CPES-$ data.
In this case we obtain a $3-\sigma$ determination of $\gamma$
(within the first quadrant) to be ${60^{+15.5}_{-19.5}}^\circ$. Using the
additional data improved the situation both by providing more statistics
and because the different data sets have different spurious solutions
leaving only the correct solution in common.

For this dash-dot curve it is instructive 
to examine the number of observables versus
the number of unknown free parameters.  First of all, for $\D^0\to
K^+\pi^-$ there is the strong phase. For each of the four parent $B^-$
decays there is a strong phase.  In the case of $B^-\to K^{*-}\D^{*0}$
there is, in addition, a parameter $R$. Again, for each of the four parent
decays there is the unknown branching ratio $b=Br(B^-\to K^{(*)-} \bar
D^{*0})$ and finally the angle $\gamma$ giving a total of 11 parameters.
On the other hand, for each combination of $B$ and $D$ decays there are
two observables, $d$ and $\bar d$ giving a total of 16 so there is an
overdetermination by 5 degrees of freedom.

As another example, consider the case where only two 
the four
combinations of $B^-\to
K^{(*)-}\D^{(*)0}$ are observed with 
$D$ decay to $K^+\pi^-$ and $CPES-$,  
then the system is still overdetermined. 
In Fig.~\ref{chisq_1} the long dashed
line takes into account only the two $B^-\to K^{-}\D^{0}$ and $B^-\to
K^{-}\D^{*0}$ and so has 6 unknown parameters determined by 8 observables.  
Clearly having some overdetermination is helpful in obtaining a good
determination of $\gamma$.

It is important to contrast thick solid curve with the long dashed
one, in  Fig.~\ref{chisq_1}. Recall both of them have $\D^0 \to
K^+ \pi^-$, $CPES-$. However, in case of the thick solid curve
the $\D^0$ originate only from $B^- \to K^- \D^0$
whereas the long dashed curve is also getting
the $\D^0$ coming from $\D^{*0} \to \D^0 + \pi^0(\gamma)$.
As a result whereas in the thick solid case there are 4 observables
and 4 unknowns for the long-dashed case its 8 observables
for 6 unknowns. That ends up making a significant difference
as is evident from the figure;
a lot more than one may naively expect just
by doubling the number of $\D^0$ or a factor of two in
luminosity. 
Infact the lower horizontal line in Fig.~\ref{chisq_1}
indicates (which corresponds to $3-\sigma$ determination
at high luminosity) that the reduction in 
error on $\gamma$ is roughly a factor of
5 (between the thick soilid and the long-dashed curves),
i.e. a saving in effective
luminosity of a factor of about 25.

It is also instructive to compare dash-dot curve, which clearly
has substantially more data, with the long-dash one. Notice that
quality of determination of $\gamma$ by the two data sets is about the
same. This suggests that once the number of observables is sufficiently large
to overdetermine the parameters, further gains by including
additional information only leads to modest gains.  

\begin{figure}   
\epsfxsize 2.9 in
\mbox{\epsfbox{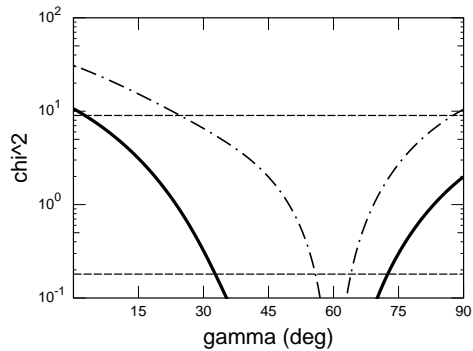}}
\caption{
The minimum value of $\chi^2$ using the four SCS modes
in Eqn.~(\ref{fourscs}) is shown with the solid line. The dashed 
dotted line shows the results with 
$D^0\to K^{*+} K^-$ and  
$D^0\to K^{*-} K^+$ together with the four parent 
$B^-\to \D^{(*)0}K^{(*)-}$ decays.   
}\label{single}
\end{figure}

We can also determine $\gamma$ by using the four SCS modes in 
Eqn.~(\ref{fourscs}). In Fig.~\ref{single} we show $\chi^2$ 
using these modes. If we again consider the four parent decays $B^-\to 
D^{(*)0}K^{(*)-}$ then we obtain the results shown with the dashed dotted 
line
which roughly gives $\gamma > 30^{\circ}$ as a $3\sigma$ bound.

%
%
%
%

%
%
%
%
%

\begin{figure}   
\epsfxsize 2.9 in
\mbox{\epsfbox{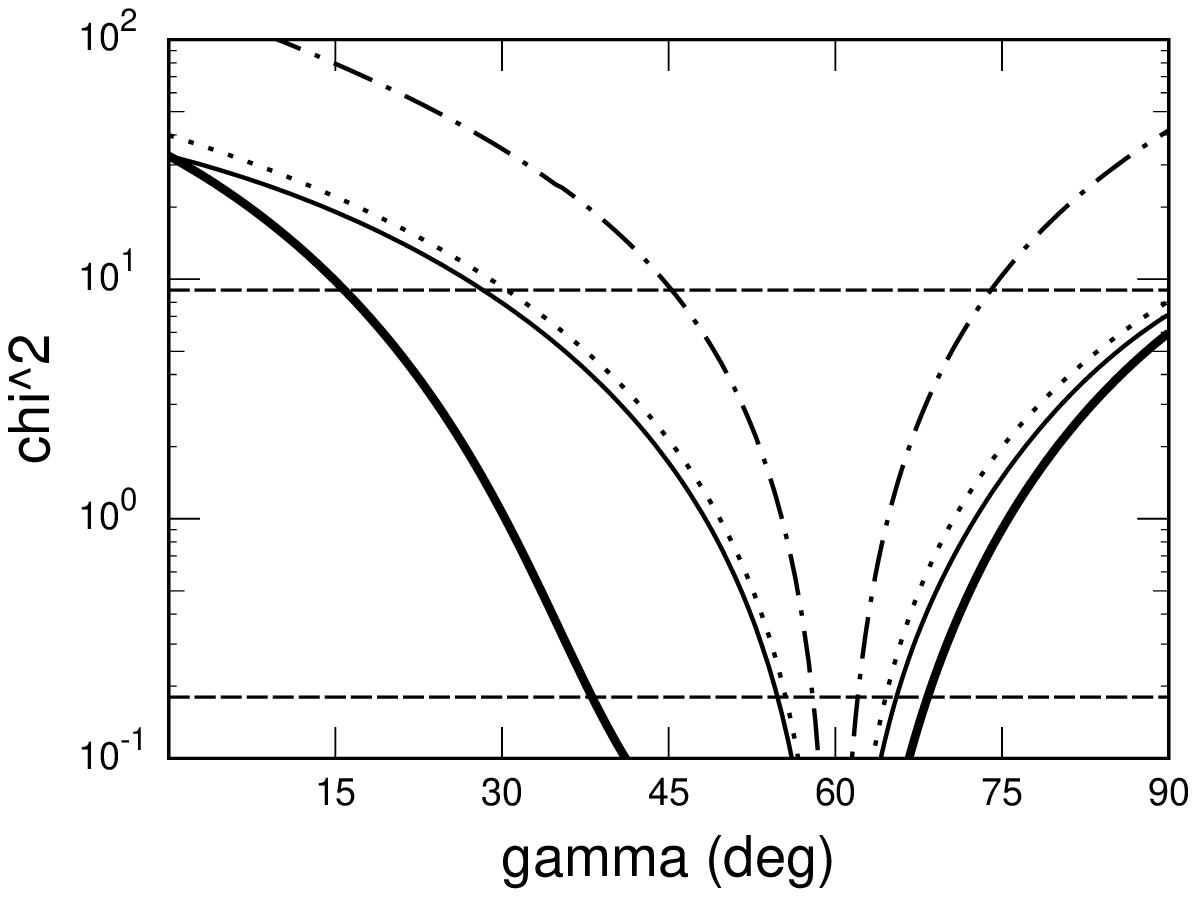}}
\caption{
The thick solid curve shows $\chi^2$ as a function of $\gamma$  taking 
both 
$B^-\to K^-[\D^0\to K^+\pi^-]$   
and
$B^-\to K^-[\D^0\to CPES-]$  
data together as in Fig.~\ref{chisq_1}.
The thin solid curve shows the result if $\zeta_{K^+\pi^-}$ 
is determined separately at a $\psi(3770)$ charm factory.
The dotted curve includes 
$B^-\to K^-[\D^0\to K^+\pi^-]$   
$B^-\to K^-[\D^0\to K^{*+}\pi^-]$   
$B^-\to K^-[\D^0\to K^+\pi^-+n\pi]$   
$B^-\to K^-[\D^0\to CPES-]$  
as well as the phase determination from a $\psi(3770)$ charm factory.
The dashed dotted curves considers all these decay modes together with
the four parent decays $B^-\to K^{*-}D^{*0}$ as well as phase determination 
from a $\psi(3770)$ charm factory.
}\label{chisq_3}
\end{figure}

Let us now consider additional sources of information which could 
constrain
these results. In the next section we will discuss the impact
of flavor tagging the $\D^0$ by means of semileptonic decays. This is
challenging but may be possible at $B$ factories. Apart from B factory
data, 
another source of additional information discussed in~\cite{adspsi}, is to
use the charm factory to determine the strong phase differences $\zeta_F$
as well as $R_F$ for inclusive modes.  In Fig~\ref{chisq_3} the thick
solid line shows the results using $B^-\to K^-D^0$ followed by $D^0$ decay
to $K^+\pi^-$ and $CPES-$ as Fig~\ref{chisq_1}. The thin solid line
shows the result if we suppose that we have the phase $\zeta_{K^+\pi^-}$
determined by a $\psi(3770)$ charm factory. Clearly this improves the
situation by removing the ambiguities.  Of course the situation can be
improved still further by including several different $D$ decay modes. In
the dotted line, we show the result where we use $\psi(3770)$ data with
the modes $K^+\pi^-$, $K^{*+}\pi^-$ and $CPES-$ and, in addition, the
inclusive decay $D^0\to K^+\pi^- + n\pi$.  Note that as well as
determining the strong phase difference for all the $D^0$ decays, the
$\psi(3770)$ data also determines the value of $R_F$ for $F= K^+\pi^- +
n\pi$. As before, we can also improve the situation by combining the data
from all $B^-\to K^{(*)-}D^{(*)0}$ combinations as is shown by the dot
dashed line.  In this case, the determination of $\gamma$ 
($3\sigma$) is $60\pm
10^\circ$.

\begin{figure}   
\epsfxsize 2.9 in
\mbox{\epsfbox{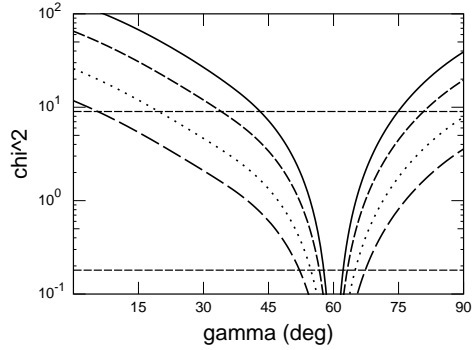}}
\caption{
The results for 
$B^-\to K^{(*)-} \D^{(*)0}$ followed by
$\D^0\to K^+\pi^-$; $CPES-$; $K^+\pi^-+n\pi$, $K^{*+}\pi^-$
where finite CP conserving backgrounds are included. The solid curve is 
for no background; the dashed curve is for signal/background=1, the 
dotted curve is for signal/background=$\frac14$ and the long dashed curve 
is for signal/background=$\frac{1}{10}$
}\label{chisq_sig_back}
\end{figure}

In the above, we have assumed that the there is no background. In general,
of course, there should be a CP-even background which will tend to
increase the error in $\gamma$.  In Fig.~(\ref{chisq_sig_back}) we show
the $\chi^2$ as a function of $\gamma$ for all four modes including all
the combinations of $B^-\to K^{(*)-} D^{(*)0}$ with the solid line. The
dashed line shows the results with a signal/background ratio of 1; the
dotted line shows the result for signal/background is $\frac14$ while the
long dashed curve shows the result for signal/background is $\frac{1}{10}$

The 3-$\sigma$ errors in $\gamma$ in these cases is
${60^{+15.1}_{-16.0}}^\circ$ for no background and 
${60^{+20.8}_{-26.1}}^\circ$ for signal/background=1.

\begin{figure}
\epsfxsize 2.9 in
\mbox{\epsfbox{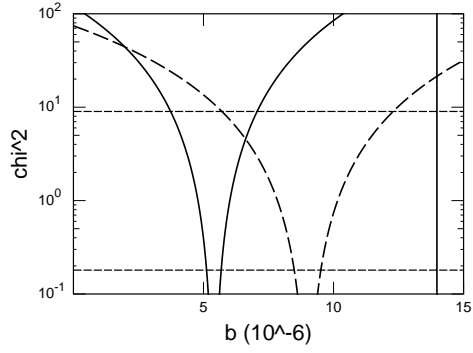}}
\caption{
The minimum value of $\chi^2$ is shown as a function of
$Br(B^-\to K^- \bar D^0)$
and
$Br(B^-\to K^{*-} \bar D^0)$ using data
for all the combinations of
$B^-\to K^{(*)-} \D^{(*)0}$ followed by
$\D^0\to K^+\pi^-$; $CPES-$; $K^+\pi^-+n\pi$, $K^{*+}\pi^-$.
The dependence on
$Br(B^-\to K^- \bar D^0)$
is indicated by the solid line
while
the dependence on
$Br(B^-\to K^{*-} \bar D^0)$
is shown by the dashed curve.
}\label{bfit_graph}
\end{figure}

Recall that the color suppressed branching ratios for
$B^- \to K^- \bar D^0$ and $B^- \to K^{*-} \bar D^0$ are 
likely to be very difficult to measure experimentally
(see however section~\ref{dtag}). The above analysis\cite{ads,ads2} 
is therefore designed to yield both these suppressed Br's
as well as $\gamma$ with the input of each data set.
Fig.~\ref{bfit_graph} serves to illustrate how well this works out 
using data
for all the combinations of
$B^-\to K^{(*)-} \D^{(*)0}$ followed by
$\D^0\to K^+\pi^-$; $CPES-$; $K^+\pi^-+n\pi$, $K^{*+}\pi^-$.
The dependence on
$Br(B^-\to K^- \bar D^0)$
is indicated by the solid line
while
the dependence on
$Br(B^-\to K^{*-} \bar D^0)$
is shown by the dashed curve.
The
3-sigma determination of $Br(B^-\to K^- \bar D^0)$ is thus $\approx \pm
1.8\times
10^{-6}$ and for $Br(B^-\to K^{*-} \bar D^0)$ it is $\approx \pm 2.25\times
10^{-6}$.

\section{Flavor Tagging $\D^0$ Mesons via Semileptonic Decays}\label{dtag}

Clearly a very useful constraint on the system of equations
Eqn.~(\ref{dsystem}) would be a direct determination of 
the branching ratio,  
$b=Br(B^-\to K^- \bar D^0)$. This can only be done if one
observes the $\bar D^0$ decay to a flavor specific semi-leptonic 
decay~\cite{ads}.
Unfortunately this is subject to a large background from semi-leptonic
decays of the parent $B^-$ since both semi-leptonic decays result in
negatively charged leptons.

The kinematics of the two kinds of processes are, however, quite different
so it may eventually be possible to determine $b$. As a case in point let
us examine the kinematics involved if one considers 
$B\to K^- \bar D^0$ followed by $\bar D^0\to \ell^-\bar \nu_\ell K^+$.

First of all, $\sum_{\ell=e,\mu}Br(\bar D^0\to \ell^-\bar \nu_\ell
K)=6.9\%$, hence the
combined branching ratio using the estimate in Table~\ref{branching_ratio_A}
is:

\begin{eqnarray}
\sum_{\ell=e,\mu}
Br(B^-\to K^-[\bar D^0\to \ell^-\bar\nu_\ell K^+)]=3.7\times 10^{-7}
\end{eqnarray}

The final state for this signal is thus $\bar \nu \ell^- K^+K^-$ and it
will be subject to the following backgrounds (including
l $=$ e and $\mu$) :

\begin{itemize}

\item[(a)] The semi-leptonic decay $B^-\to \bar \nu \ell^- [D^0\to
K^+K^-]$.
This has a combined branching ratio of $1.7\times 10^{-4}$.

\item[(b)] The semi-leptonic decay $B^-\to \bar \nu \ell^- [D^0\to
\pi^+K^-]$ where the $\pi^+$ is misidentified as a $K^+$. The combined
branching ratio is $1.6\times 10^{-3}$. Of course this must be multiplied
by the rate of mis-identification for a given state.

\item[(c)] 
$B^-\to \bar \nu \ell^-  + [X_c\to K^+K^-+X]$,
with a branching ratio $\approx (10^{-2} - 10^{-3})$.

\item[(d)] $B^-\to \bar \nu \ell^- K^+ K^- $; the branching ratio is
crudely estimated to be around $3 \times 10^{-5}$. 

\item[(e)] Backgrounds form continuum events

\end{itemize}

Against these backgrounds one can apply the following kinematic cuts.

\begin{itemize}

\item[(1)] For the signal the energy of the $K^-$ in the rest frame of the
$B^-$ is fixed to be $E_{K^-}=(m_B^2+m_K^2-m_D^2)/(2m_B)$.

\item[(2)] The missing neutrino leads to the constraint
$|p_B-p_{K^-}-p_{\ell}-p_{K^+}|=0$.

\item[(3)] For backgrounds of the type (a), the invariant mass of the $K^+
K^-$
system will be $m_D$; indeed, to eliminate the related backgrounds from
$D^0\to K^+K^-+X$ one may use the cut $|p_{K^+}+p_{K^-}| > m_D$.

\item[(4)] Signal events will have three distinct vertices while continuum
background will have only 1.

\end{itemize}

Background (a) and (b) are $\sim 10^3 \times$ signal. 
Cut (3) would remove these entirely
except for momentum resolution but the additional cuts (1) and (2) which are
satisfied by the signal may be sufficient to control this large
background. Backgrounds (d) is probably $O(10^2)$ times the
signal; (c) is probably even a lot bigger.
(d) in particular would pass cuts (2) and (3) and would have to be
reduced by cuts (1) and (4). Likewise 
background (c) would have to be reduced by the
monoenergetic kaon cut (1). Background (e) could, in principle, also be
large and it is not clear whether cuts (1) and 
(4) together with the standard cuts
against the continuum can be sufficient to eliminate it.

In spite of the difficulty in determining $b$, in the case where the $D^0$
decays to a CP eigenstate, a relatively weak bound, 
on b, can be helpful in a
determination of $\gamma$.

\begin{figure}   
\epsfxsize 2.9 in
\mbox{\epsfbox{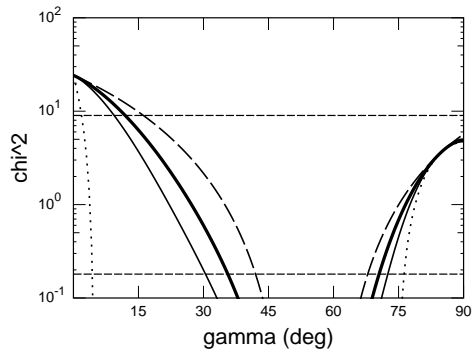}}
\caption{
$\chi^2$ as a function of $\gamma$ for $B^-\to K^-[D^0\to CPES-]$. The 
dotted curve is the same as in Fig.~\ref{chisq_1}, the long dashed curve 
assumes that $b$ has a 10\%  1-$\sigma$ error, the thick line assumes 
a 100\% error 
and the thin solid line assumes a 200\% error.
}\label{glw_bound}
\end{figure}

In Fig.~\ref{glw_bound} we show the $\chi^2$ curve obtained just from
the GLW modes, 
$B^-\to K^-[D^0\to CPES-]$, as in Fig.~(\ref{chisq_1}) with the dotted
line. The dashed line, the thick solid line and the thin solid line
correspond to the result obtained assuming a 10\%, 100\% and 200\%
1-$\sigma$ Gaussian error in the determination of $b$ respectively. These
lead to 3-$\sigma$ lower bounds on $\gamma$ of $25.2^\circ$, $16.8^\circ$
and $12.6^\circ$ with the prospect of improvement at higher statistics.
The reason for this is that the CP violation in this case is relatively
small $\sim 15\%$ so that the solutions with small $\gamma$ correspond to
a situation where $b$ is large but $\zeta_B\sim 180^\circ$ so there is
near cancellation between the two channels. This is quite different from
the actual situation where $b<<a$ so a modest bound on $b$ improves the
situation greatly.

If an experimental value for $b$ is not available, a theoretical estimate
can serve at the expense of the result becoming model dependent. From the
above, such an estimate need not be very precise to be of some utility. On
the other hand, the decay $B^-\to K^- \bar D^0$ is color suppressed making
reliable theoretical predictions more difficult.

\begin{figure}   
\epsfxsize 2.9 in
\mbox{\epsfbox{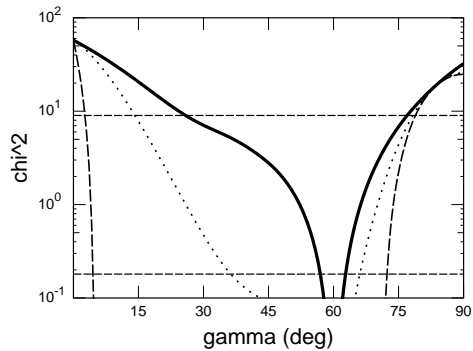}}
\caption{
The curves are as in Fig.~\ref{glw_bound} except all four $B^-$ decays 
of the form $B^-\to K^{(*)-}D^{*(0)}$ are considered. 
}\label{glw_bound_allstar}
\end{figure}

In Fig.~\ref{glw_bound_allstar} we show the analogous calculation where
we have combined the data from all the combinations of $B^{-}\to K^{(*)-}
[D^{(*)0}\to CPES-]$ with a determination of $b$ to 10\% (thick solid
line), 200\% (dotted line) and unconstrained (dashed line). These give the
corresponding 3-sigma lower bound on $\gamma$ of $40.2^\circ$,
$16.2^\circ$ and $3^\circ$ respectively and upper bounds of 
$76.8^\circ$,
$78.0^\circ$ and $78.6^\circ$, respectively.

%
%
%
%


\section{Impact of $D\bar D$ Mixing}

As has been discussed previously in the literature~\cite{ss1,ads2}, 
the effects of $D\bar
D$ mixing could be significant on some of the combined branching
ratios. This is particularly true for final states such as $K^+ \pi^-$
because the favored decay $B^-\to K^- D^0$ could be followed by the
favored $\bar D^0\to K^+\pi^-$ if the $D^0$ were to oscillate to a $\bar
D^0$. If the probability of oscillation is $O(1\%)$ then this channel
might be comparable to the direct DCS rate for $B^-\to K^-[D^0\to
K^+\pi^-]$ which is assumed to be the only channel in the absence of
oscillation.

As discussed in the appendix, mixing is controlled by 4 parameters, $x$,
$y$, $A$ and $\phi$. If CP is conserved in $D$ decays,
which is an excellent approximation in the SM,
then $A=\phi=0$. The
standard model predicts that the parameter controlling
the mass difference, $x\sim
O(10^{-4})$. The Standard Model prediction for the parameter 
relevant for the life time difference,  
$y$, is less certain since it could be enhanced by final state
hadronic interactions. It has been suggested~\cite{gronau_d} that $y$ could
be as large as $10^{-2}$.

\begin{figure}   
\epsfxsize 2.9 in
\mbox{\epsfbox{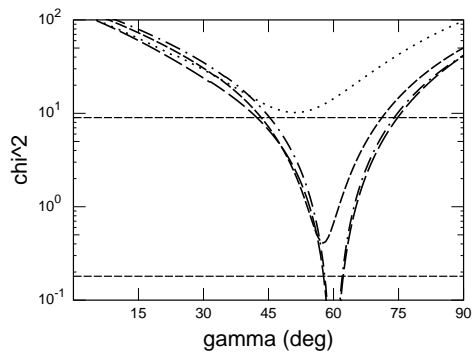}}
\caption{
The solid line shows the fit to the data from $K^+\pi^-$, $K^+\pi^-+n\pi$,
$CPES-$ and $K^{*+}\pi^-$ with no $D$ mixing. the dashed line shows the
fit, assuming $y=0$ to the data in the case $y=0.01$
the dotted line shows the
fit, assuming $y=0$ to the data in the case $y=0.1$
while the dash dotted line shows fit where a non-zero value of $y$ is
taken into account where the data is generated with $y=0.1$.
}\label{chisq_mix1}
\end{figure}

%
%
%
%
%
%

%
%

In Fig.~\ref{chisq_mix1} 
the solid line is a fit to the data for
$K^+\pi^-$, $K^+\pi^-+n\pi$, $CPES-$ and $K^{*+}\pi^-$ final states as
discussed before. The dotted line shows the result if $y=0.01$ but the
result is fit assuming there is no mixing.  The minimum value of $\chi^2$
has been shifted down to $58^\circ$. The dashed line shows the same
calculation for $y=0.05$ and the minimum value of $\chi^2$ is shifted down
to $51^\circ$. The dashed dotted line shows the $\chi^2$ for $y=0.05$
where the fit is done taking into account a non-zero value of $y$. In this
case the fit value of $\gamma$ is ${60_{-16}^{+15.5}}^\circ$. 

%
%

Of course data from a $\psi(3770)$ charm factory should be
able to improve significantly on the 
constraints on the $D^0\bar D^0$ mixing parameters
and thereby it could immensely aid in the accurate determination
of $\gamma$ from the $B \to K D^0$ method.

\section{Self Tagging Neutral $B$ Decays}

It is also useful to add information obtainable from the decay of neutral
$B$'s. Since $B^0$ undergoes oscillation there are two strategies which
can be considered:

\begin{itemize}
\item 
The $B^0$ decays which are flavor non-specific such as $B^0\to K_S D^0$ or
$B^0\to K^{*0} D^0$ where $K^{*0}\to K_S\pi^0$. In this case the
oscillation of the $B^0$ plays a role.
\item Self-tagging~\cite{desh,dunietz_selftag}
decays which are flavor specific such as $B^0\to \bar
K^{*0} D^0$ where $K^{*0}\to K^-\pi^+$. In this case the oscillation of
the $B^0$ does not play a role.
\end{itemize}

In the flavor non-specific case, the oscillation phase $\beta$ in the Standard
Model 
enters in such a way that 
the main dependence is on
$\delta \equiv 2\beta+\gamma \equiv \beta -\alpha +\pi$. 
This has been discussed
in~\cite{deltapaper,sanda,kayser, robf}.


Here we will discuss the former case where the oscillation plays no role
and so only direct CP violation via flavor specific decays is used in
a determination of $\gamma$; for instance, via the decay $\bar
B^0\to [K^{*0}\to K^-\pi^+]\D^0$.  The main difference with the charged
($B^{\pm}$) decay is that the $b\to c$ channel is color suppressed and
thus will be about 9 times smaller than the charged case. Thus we will
estimate the branching ratios as shown in Table~\ref{branching_ratio_A}.
The $b\to u$ channel proceeds via the same diagram in the charged and
neutral cases hence the branching ratios in these cases should be about
the same as shown in Table~\ref{branching_ratio_A}.

Because of the suppression of the $b\to c$ channel, if $D^0$ decays to a
DCS mode the amplitudes no longer are matched and so the CP violation will
be somewhat smaller than in the charged case. Conversely the CP violation
in modes with 
$D$ decays to $CPES$ will now be larger.

%
%
%
%
%
%

\begin{figure}   
\epsfxsize 2.9 in
\mbox{\epsfbox{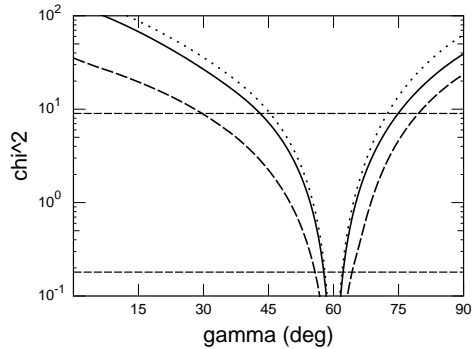}}
\caption{
The solid curve 
is generated using charged $B^-$ data and 
is the same as the solid curve in 
Fig.~\ref{chisq_sig_back}
The dashed curve uses only neutral data with the two parent decays
$\bar B^0\to [\bar K^{*0}\to K^-\pi^0]\D^{(*)0}]$
and subsequent $\D$ decay to $K^+\pi^-$,
$K^{*+}\pi^-$, $K^+\pi^-+n\pi$, 
$CPES-$. The dotted curve combines the data set used in the solid curve 
(charged B) 
with the data set used in the dashed curve (neutral B). 
}\label{neutral_fig_1}
\end{figure}

In Fig.~\ref{neutral_fig_1} we illustrate the use of such modes. The 
solid line, for reference is the same as in Fig.~\ref{chisq_sig_back}, 
including data from all combinations of $B^-\to K^{(*)-}\D^{(*)0}$ with 
subsequent $\D$ decay to $K^+\pi^-$, $K^{*+}\pi^-$, $K^+\pi^-+n\pi$, 
$CPES-$. The dashed curve considers the neutral $B$ decays 
$\bar B^0\to [\bar K^{*0}\to K^-\pi^0]$ with subsequent $\D$ decay to the 
same four final states. Because of the extra color suppression, it is not 
surprising that it falls below the solid line. Of course we can combine 
the two data sets since the strong phases of the $\D$ decays should be 
common to each in which case one obtains the dotted curve improving the 
situation slightly
over the solid curve.

\section{Multibody $B$ Decays}\label{multibody} 

Two body decay modes of the B to $\D^0$ suffer from the problem that at
least one of the channels is color suppressed.  A method to circumvent
this has been suggested by Aleksan, Petersen and Soffer~(APS)~\cite{bkdpi}
that decays of the form $B^-\to K^-\pi^0 D^0$ may be used to extract
$\gamma$. On the quark level the processes are the same as the two body
$B$ decays considered above but in this manifestation there are two
important differences:

\begin{enumerate}
\item Both the $b\to c$ and $b\to u$ channels are color allowed.
\item The amplitude is a function of the 3-body phase space.
\end{enumerate}

The fact that the $b\to u$ channel is color allowed is important since it 
can improve the statistics, particularly if $D$ decays to CPES. To take 
full advantage of this, however, we must overcome 
the fact that this is an inclusive state spread over phase space (i.e. the
Dalitz Plot).

The APS method~\cite{bkdpi} is to extract from the data a number of
different amplitudes and their relative phases by fitting to the Dalitz
plot using a model with resonance and continuum terms and comparing the 
$D^0$ decay to CPES and flavor specific modes. 

Here we will consider a complimentary approach where we integrate over the
Dalitz plot using the formalism of Eqn.~(\ref{rdef_2}) in the Appendix 
where
there is an additional coherence parameter associated with the $B$
decay. 
In this approach, though, 
several different three or more body decays may
be taken together which will tend to increase the statistics.

First, let us compare the number of degrees of freedom versus the number
of observables in such a case. Suppose that we consider $N$ inclusive
decays of the form $B\to \D^0 K+X_i$ ($i=1\dots N$) where the $\D^0$
subsequently decays into $CPES-$, $M_E$ exclusive modes and $M_I$
inclusive modes.  For each $B$ decay there are 3 parameters, $b$, $\zeta$
and $R_F$. For each inclusive $D$ mode there is one phase while for each
inclusive D mode there is a phase and a coherence factor. In addition, of
course, $\gamma$ is also unknown giving $3N+M_E+2M_I+1$ parameters.  On
the other hand, for each combination of B decay modes with each $\D$ decay
mode there are two observables, $d$ and $\bar d$ giving a total of
$2N(M_E+M_I+1)$ observables. Thus, $\gamma$ may be determined if

\begin{eqnarray}
2N(M_E+M_I+1) \geq 3N+M_E+2M_I+1
\label{df_balance}
\end{eqnarray}

\noindent 
where for the case of 
equality applied there would likely be ambiguities. If we 
rearrange the above equation assuming that $M_E+M_I>0$ 
we obtain:

\begin{eqnarray}
N\geq
\frac{M_E+2M_I+1}{2M_E+2M_I-1}
\label{n_bound}
\end{eqnarray}

\noindent 
From this form we see that the lower bound on $N$ is 3 only in
the case of $M_E=0$ and $M_I=1$. In other cases the lower bound is either
1 or 2 (taking into account that $N$ is an integer). Thus at least 1, 2 or
3 different $B$ decays must be observed depending (according to
eqn.~(\ref{n_bound}))  on how many different $D$ decays are considered
assuming, in all cases, that we observe the CPES- decays of the $\D^0$.
 
\begin{figure}   
\epsfxsize 2.9 in
\mbox{\epsfbox{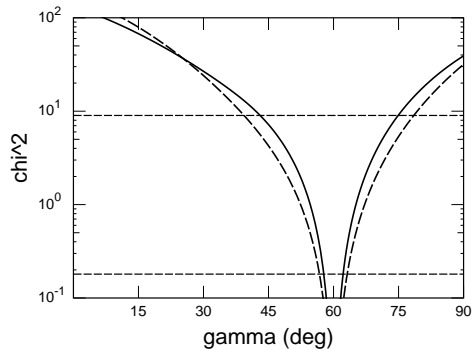}}
\caption{
The solid curve is the same as in Fig.~\ref{chisq_sig_back}
while the dashed one is the $\chi^2$ for 
$\gamma$ determined by 
the three body $B^-$ decay modes
$\D^0 K^- \pi^0$,
$\D^{*0} K^- \pi^0$ and
$\D^0 K^{*-} \pi^0$
using the assumptions discussed in the text.
}\label{three_bdy_1}
\end{figure}

As an illustration of this let us assume that the three inclusive modes
$B^-\to \D^0 K^- \pi^0$, $B^-\to \D^{*0} K^- \pi^0$ and $B^-\to \D^0 K^{*-}
\pi^0$ are observed with $\D^0\to CPES-$, $K^+\pi^-$ and $K^{*+}\pi^-$ so
that $N=3$, $M_E=2$ and $M_I=0$. Clearly Eqn.~(\ref{df_balance}) is
satisfied since there are 18 observables for 12 unknowns.  To carry out
our calculation we will assume that
$a(D^{(*)0}K^{(*)-}\pi^0)=a(D^{(*)0}K^{(*)-})$ and $b(\bar
D^{(*)0}K^{*-}\pi^0)=9b(\bar D^{(*)0}K^{(*)-})$ since it is not color
suppressed; also we take $R=0.5$ in each case. The results are shown with the
dashed line in Fig.~\ref{three_bdy_1}.

\section{Prospects for $\gamma$ at a Super-B Factory}

Improvements at the existing asymmetric B-factories at KEK
and SLAC are expected to allow luminosity increase to
around $10^{35} cm^{-2}sec^{-1}$. 
Further increase in luminosity 
to $10^{36}cm^{-2}sec^{-1}$ and beyond will most likely
require a new machine\cite{my_izu}.
The asymmetric B-factories at KEK and SLAC have made
a rather accurate determination of $\sin(2\beta)$
and established that the CKM-phase is the dominant contributor
to the observed CP asymmetry in $B \to \psi K_s$,
which has to be regarded an important milestone
in our understanding of CP violation phenomena.
There is now considerable interest at the construction
of an asymmetric ``Super-B Factory"(SBF) with luminosity
$10^{36}cm^{-2}sec^{-1}$ or more\cite{izu,slac}.  
Precision determinations of all the angles of the unitarity triangle,
roughly in the range of the intrinsic theory 
error for each of the three angles, 
and the search for beyond the SM source(s) of CP violation,
which are bound to exist, 
constitute important motivations for such an effort. 

In particular, $\gamma$ can be determined very cleanly from
direct CP violation via the generic $B \to K(^*) D^0(^*)$ processes. There are
multitude of available strategies and modes; several of them we
dealt in this paper. Using Fig~\ref{chisq_1} as a guide
we can anticipate possible determination of $\gamma$ at a
Super-B factory with a 3-$\sigma$ error of a few 
($\approx 2$) degrees.

Admittedly backgrounds could make things difficult
in an actual experimental setup, as Fig~\ref{chisq_sig_back}
indicates; from the figure we see that 
the accuracy in determination of $\gamma$ from these methods
may, due to backgrounds, get degraded to $\approx 7^\circ$
at the Super-B Factory. It should be noted though that 
strategies used in
Fig~\ref{chisq_1} and
Fig~\ref{chisq_sig_back}
are far from exhaustive. For example,
one promising technique which we had suggested earlier \cite{ads,ads2} 
is Dalitz plot study of 3-body decays of $D^0$.
Since some experimental data had existed at that time for
$D^0 \to K^+ \pi^- \pi^0$ from Fermilab experiment E687 
\cite{E687}, we had used that specific mode to
illustrate the method.  More recently CLEO collaboration\cite{cleo_dalitz}
has also studied $D^0 \to K_s \pi^+ \pi^-$ which is another
interesting candidate where the analysis \cite{ads2}
could be readily
applied to extract $\gamma$. 
Indeed preliminary studies of this mode by the BELLE collaoboration
\cite{belle_3} are quite encouraging.

In addition to Dalitz plot studies of $D^0$ decays, charm factory
running on $\psi(3770)$ could be very helpful in determination
of $\gamma$ from $B \to K D$ approaches as Fig~\ref{chisq_3} 
illustrates; see also section~\ref{multibody} and \cite{adspsi}.   


\section{Summary and Outlook}

In this paper we have 
extended our previous studies\cite{ads,ads2,adspsi}
on extraction of $\gamma$
using direct CP violation in $B \to K D$ processes.
In principle, these methods are theoretically very clean.
The irreducible theory error originating from
higher-order weak interactions is $O(10^{-3})$\cite{pristine},
i.e. in all likelihood even smaller 
than the theory error in deducing the angle 
$\beta$ using time dependent CP
asymmetry in $B \to \psi K_s$.  
However, $\gamma$ determination 
from $B \to K D$ is much harder 
than $\beta$ from $B \to \psi K_s$. 

This study strongly suggests that
the demands on machine luminosity can be significantly alleviated
if a combination of strategies is used.
One interesting handle that we examined here
which looks rather promising is to include
$D^{*0}$ from $B \to K^{(*)}D^{*0}$. The formalism for
the use of $D^0$ decays is identical to
$B \to K^{*} \D^0$ after $D^{*0} \to
\pi^0(\gamma) + D^0$. 

Similarly including $K^*$ 
(via e.g. $B^- \to K^{*-} D^0$) along with $B^-
\to K^- D^0$ is helpful. 
Also it of course helps a great deal to use both
CP eigenstates\cite{glw} along with CP non-eigenstates of $D^0$,
whether they be doubly Cabibbo suppressed\cite{ads,ads2} 
or singly Cabibbo suppressed\cite{scs}.  

While flavor tagging of the $D^0$ and $\bar D^0$ is likely
to be a challenge,  we emphasized that kinematics and topology
of the signal
events is quite distinct from that of many of the backgrounds
so that semi-leptonic tags may have a chance at least
for a relatively imprecise determination
of the needed color-suppressed branching ratio.
If such a determination could be made, say at super-B factory with
an error on the branching ratio of around 10-20\%, it could become helpful
in $\gamma$ determination.   

Many of the modes relevant for gamma determination
should exhibit large direct CP-asymmetries; this is especially
so in the case of doubly Cabibbo suppressed modes of $D^0$. 
However, as Tables~(\ref{asy_tab},\ref{asy_tabB},\ref{asy_tabBx})
illustrate CP asymmetries 
in many other modes are also appreciable.  
Indeed, at least in the initial stages, it may even be useful
to target the search for such large 
asymmetries.   

It is difficult to overemphasize the important role
that a charm factory running at $\psi(3770)$ can play
in the extraction of $\gamma$. Charm factory can of course help
by improving on the mixing parameters of the
$D^0- \bar D^0$ complex. It can also help in determination
of the doubly Cabibbo suppressed branching ratios
as well as in the needed strong phases. Quantum entanglement
in $\psi \to D^0 \bar D^0$ decays 
can also be exploited to determine the coherence factor
and the strong phase for 
inclusive $D^0$ decays\cite{adspsi}. 

Note also that all of these studies of charm physics
at the charm factory not only have important application
to extraction of $\gamma$ from direct CP studies
in $B \to K D$ processes, 
precisely the same information on D-decays
and D-mixing can also be used 
in time dependent CP studies
in $B^0 \to K^0 D^0(^*)$; in that case one gets
the linear combination of unitarity angles $\delta \equiv
(2 \beta + \gamma)
\equiv (\beta - \alpha + \pi)$ and furthermore that  method
also gives $\beta$\cite{kayser,deltapaper} providing an additional
test of the CKM-paradigm.  

While B-factories with about $10^9$ B-pairs are likely to be able to
make appreciable progress in determination of $\gamma$, super-B 
factory with luminosity $ \ge 10^{10}$ B-pairs will be needed
to extract $\gamma$ with an accuracy roughly commensurate with
the intrinsic theory error that these methods allow. This 
in itself should
constitute an important goal of B-physics in general and super-B
factory in particular.   

\section{Acknowledgments}

We would like to thank Tom Browder, Riccardo Faccini,
Tim Gershon, Masashi Hazumi, Gerald Raven, Abi Soffer, Sheldon
Stone and Sanjay Swain.  This research was supported by Contract Nos.\
DE-FG02-94ER40817 and DE-AC02-98CH10886.

\section*{Appendix: Formalism}

Let us now present a summary of the formalism for determining 
the rates for $B^{-}\to K^{(*)-}[D^{(*)0}\to F]$ and related processes. 
Here we will discuss the rates in four situations: 

\begin{enumerate}
\item
$F$ is an exclusive state; without $D^0\bar D^0$ oscillations.
\item
$F$ is an inclusive set of states; without $D^0\bar D^0$ oscillations.
\item
$F$ is an exclusive state
where
$D^0\bar D^0$ oscillations are considered
\item
$F$ is an inclusive state
where
$D^0\bar D^0$ oscillations are considered.
\end{enumerate}

For this purpose exclusive state refers to a 
state which is governed by a single
quantum amplitude, for instance $D^0\to K^+\pi^-$. In contrast an
inclusive state is composed of a decay where there are a number of
amplitudes contributing incoherently. Some examples are inclusive states
made of states with different particle content, for instance $K^++n\pi$ or
multibody final states integrated over phase space, for instance
$K^+\pi^+\pi^0$ integrated over the Dalitz plot. Note that each point in
the $K^+\pi^-\pi^0$ Dalitz plot is an exclusive state. 

First let us take $f$ to be an exclusive state. Let us denote by $d$ the
branching ratio for $B^-\to K^{(*)-}[D^{(*)0}\to f]$ where $D^0$
means a general mixture of $D^0$ and $\bar D^0$ and $\bar d$ the charge
conjugate branching ratio for $B^+\to K^{(*)+}[D^{(*)0}\to \bar f]$.

For this decay let $a$ be the branching ratio for $B^-\to K^- D^0$ and $b$
be the branching ratio for $B^-\to K^- \bar D^0$.
Likewise let us denote by $c$ the branching ratio for $D^0\to f$ and
$\bar c$ the branching ratio for $\bar D^0\to f$. Assuming that $D^0$
mixing is negligible\cite{ads,ads2}, 

\begin{eqnarray}
d&=& a c_f + b\bar c_f + 2\sqrt{abc_f\bar c_f}\cos(\zeta_B+\zeta_f+\gamma)
\nonumber\\
\bar d&=& a c_f + b\bar c_f + 2\sqrt{abc_f\bar
c_f}\cos(\zeta_B+\zeta_f-\gamma)
\label{exclusive_nomix}
\end{eqnarray}

\noindent where $\zeta_B$ is the strong phase difference between $B^-\to
K^{(*)-}D^{(*)0}$ and $B^-\to K^{(*)-}\bar D^{(*)0}$ and $\zeta_f$ the
strong phase difference between $D^0\to f$ and $D^0\to \bar f$.

Let us now consider the generalization to the case where $F$ is an 
inclusive state, $F=\{f_1,\dots,f_n\}$ and denote: 

\begin{eqnarray}
R_F e^{i\zeta_F}=
\frac{\sum_i \sqrt{c_{f_i}\bar c_{f_i}}e^{i\zeta_{f_i}}}
{\sqrt{ c_F \bar  c_F}}
\label{rdef_1}
\end{eqnarray}

where $c_F = \sum_i c_{f_i}$ and similarly $\bar c_F$. 
For the decay to an inclusive final state $F$, then, we can generalize
eqn.~(\ref{exclusive_nomix}) to:

\begin{eqnarray}
d&=& a c_F + b\bar c_F 
\nonumber\\
&&
+ 2R_F \sqrt{abc_F\bar
c_F}\cos(\zeta_B+\zeta_F+\gamma)
\nonumber\\
\bar d&=& a c_F + b\bar c_F 
\nonumber\\
&&
+ 2R_F \sqrt{abc_F\bar
c_F}\cos(\zeta_B+\zeta_F-\gamma)
\label{inclusive_nomix}
\end{eqnarray}

\noindent
Note that in the case where $F$ is an exclusive state $R_F=1$.

Another, related generalization is the case where the initial $B^-$ decays
to an inclusive state. For instance this is the case in $B^-\to
K^{*-}D^{*0}$ because there are three helicity amplitudes. Of course the
amplitudes may be separated through the consideration of angular
distributions in the vector decays. Another instance is a three body decay
such as $B^-\to K\pi D^0$.

In general, we can consider $B^-\to G$ where $G$ is an inclusive set of
states. Thus, each $g_i$ contains one (generic) $D^0$ meson. We will use
$g_i^\prime$ to denote, specifically, the version of the state where the
neutral $D$ is in a $|D^0>$ state and $g_i^{\prime\prime}$ to be the case
where the neutral $D$ is in a $|\bar D ^0>$ state. Let us denote the
branching ratios $B^-\to g_i^\prime$ by $a_{g_i}$ and $B^-\to
g_i^{\prime\prime}$ by $b_{g_i}$.
One could, in principle, determine $a_{g_i}$ and $b_{g_i}$ by observing
the semileptonic decay of the $D^0$.

Thus, 
in analogy to Eqn.~(\ref{rdef_1}) let us define:

\begin{eqnarray}
R_G e^{i\zeta_G}=
\frac{
\sum_i \sqrt{a_{g_i}\bar b_{g_i}}e^{i\zeta_{g_i}}
}{ \sqrt{ a_{g_i}  b_{g_i} }
}
\label{rdef_2}
\end{eqnarray}

\noindent
If the neutral $D$ in $G$ subsequently decays to an inclusive state $F$,
then the combined branching ratio is:

\begin{eqnarray}
d&=& a_G c_F + b_G \bar c_F 
\nonumber\\
&&
+ 2R_F R_G \sqrt{a_Gb_Gc_F\bar
c_F}\cos(\zeta_G+\zeta_F+\gamma)
\nonumber\\
\bar d&=& a_G c_F + b_G \bar c_F 
\nonumber\\
&&
+ 2R_F R_G \sqrt{a_Gb_Gc_F\bar
c_F}\cos(\zeta_G+\zeta_F-\gamma)
\label{inclusive2_nomix}
\end{eqnarray}

Let us now consider the case where $D^0\bar D^0$ mixing is present. 

First, let us consider the fully general case where, following the usual
formalism~\cite{mix_formal,ads2}, the eigenstates of neutral $D$ meson are:

\begin{eqnarray}
|D_1\ket&=&p|D^0\ket+q|\bar D^0\ket \nonumber\\
|D_2\ket&=& p|D^0\ket-q|\bar D^0\ket
\end{eqnarray}

\nonumber
with corresponding masses $m_1$, $m_2$ and widths $\Gamma_1$,
$\Gamma_2$. These are characterized by the parameters:

\begin{eqnarray}
x=\frac{m_2-m_1}{ \Gamma},\ \ \ y=\frac{\Gamma_2-\Gamma_1}{ 2\Gamma}
\end{eqnarray}

\noindent
where $2\Gamma=\Gamma_1+\Gamma_2$.

If $p/q\neq 1$ then CP is violated in the mixing. If this is indeed the
case, one would expect that there would also be CP violating phases in the
decay. In the following we will assume that $D$ decays are CP conserving
and then indicate how to generalize to the case of CP violation in the
final state. 
In this case the decay amplitudes of $D^0$ only have strong phases and we
denote:

\begin{eqnarray}
\frac{p}{q}=exp^{A+i\phi}
\end{eqnarray}

Thus, 
generalizing 
Eqn.~(\ref{inclusive2_nomix}) 
to the case where 
there is $D^0$ mixing and integrating over time:

\begin{eqnarray}
d&=& {Q+P\over 2}
\bigg [
a_G c_F+b_G\bar c_F 
\nonumber\\&&
+2 R_FR_G\sqrt{a_G c_Fb_G\bar c_F}\cos(\eta_G+\gamma+\eta_F) 
\bigg ]
\nonumber
\end{eqnarray}

\begin{eqnarray}
&&+
{Q-P\over 2}
\bigg [
e^{2A}a_G c_F+e^{-2A}b_G \bar c_F
\nonumber\\&&
+2 R_FR_G\sqrt{a_G c_Fb_G\bar c_F}\cos(\eta_G+\gamma-\eta_F) 
\bigg ]
\nonumber
\end{eqnarray}

\begin{eqnarray}
&&
-yQ
\bigg [
c_F\sqrt{a_Gb_G}R_Ge^{A}\cos(\eta_G+\gamma+\phi)
\nonumber\\&&
+
\bar c_F\sqrt{a_Gb_G}R_Ge^{-A}\cos(\eta_G+\gamma-\phi)
\nonumber\\&&
+
a_G\sqrt{c_F\bar c_F} e^{-A} R_F \cos(\eta_F+\phi)
\nonumber\\&&
+
b_G\sqrt{c_F\bar c_F} e^{A} R_F \cos(\eta_F-\phi)
\bigg ]
\nonumber
\end{eqnarray}

\begin{eqnarray}
&&
+xP
\bigg [
c_F\sqrt{a_Gb_G}R_Ge^{A}\sin(\eta_G+\gamma+\phi)
\nonumber\\&&
+
\bar c_F\sqrt{a_Gb_G}R_Ge^{-A}\sin(\eta_G+\gamma-\phi)
\nonumber\\&&
+
a_G\sqrt{c_F\bar c_F} e^{-A} R_F \sin(\eta_F+\phi)
\nonumber\\&&
+
b_G\sqrt{c_F\bar c_F} e^{A} R_F \sin(\eta_F-\phi)
\bigg ]
\end{eqnarray}

\begin{eqnarray}
\bar d&=& {Q+P\over 2}
\bigg [
a_G c_F+b_G\bar c_F 
\nonumber\\&&
+2 R_FR_G\sqrt{a_G c_Fb_G\bar c_F}\cos(\eta_G-\gamma+\eta_F) 
\bigg ]
\nonumber
\end{eqnarray}

\begin{eqnarray}
&&+
{Q-P\over 2}
\bigg [
e^{-2A}a_G c_F+e^{2A}b_G \bar c_F
\nonumber\\&&
+2 R_FR_G\sqrt{a_G c_Fb_G\bar c_F}\cos(\eta_G-\gamma-\eta_F) 
\bigg ]
\nonumber
\end{eqnarray}

\begin{eqnarray}
&&
-yQ
\bigg [
c_F\sqrt{a_Gb_G}R_Ge^{-A}\cos(\eta_G-\gamma-\phi)
\nonumber\\&&
+
\bar c_F\sqrt{a_Gb_G}R_Ge^{A}\cos(\eta_G-\gamma+\phi)
\nonumber\\&&
+
a_G\sqrt{c_F\bar c_F} e^{A} R_F \cos(\eta_F-\phi)
\nonumber\\&&
+
b_G\sqrt{c_F\bar c_F} e^{-A} R_F \cos(\eta_F+\phi)
\bigg ]
\nonumber
\end{eqnarray}

\begin{eqnarray}
&&
+xP
\bigg [
c_F\sqrt{a_Gb_G}R_Ge^{-A}\sin(\eta_G-\gamma-\phi)
\nonumber\\&&
+
\bar c_F\sqrt{a_Gb_G}R_Ge^{A}\sin(\eta_G-\gamma+\phi)
\nonumber\\&&
+
a_G\sqrt{c_F\bar c_F} e^{A} R_F \sin(\eta_F-\phi)
\nonumber\\&&
+
b_G\sqrt{c_F\bar c_F} e^{-A} R_F \sin(\eta_F+\phi)
\bigg ]
\end{eqnarray}

\noindent
where
\begin{eqnarray}
P={1\over 1+x^2},\ \ \ \ Q={1\over 1-y^2}
\end{eqnarray}

Note that in the limit that $x$, $y$, $A$, $\phi\to 0$ these
equations reduce to 
Eqn.~(\ref{inclusive2_nomix}).


\begin{thebibliography}{99}


\bibitem{ckm}
N. Cabibbo, Phys.\ Rev.\ Lett.\ {\bf10}, 531 (1963); M.
Kobayashi and T. Maskawa, Prog.\ Th.\ Phys.\ {\bf49}, 652 (1973). 




\bibitem{belle_beta} K.~Abe et al (BELLE Collab), Phys.\ Rev.\ D {\bf
66}, 071102 (2002)



\bibitem{babar_beta} B.~Aubert et al (BABAR Collab), Phys.\ Rev.\ Lett.\ {\bf
89}, 201802 (2002).




\bibitem{glw}M.~Gronau and D.~Wyler, Phys.\ Lett.\ {\bf B265}
(1991); M.~Gronau and D.~London., Phys.\ Lett.\ {\bf B253}, 483 (1991).




\bibitem{ads}
D.~Atwood, I.~Dunietz and A.~Soni,
Phys.\ Rev.\ Lett.\  {\bf 78}, 3257 (1997).


\bibitem{ads2}  D.~Atwood, I.~Dunietz and A.~Soni, 
Phys.\ Rev.\ D {\bf 63}, 036005 (2001).


%
%
\bibitem{adspsi}
D.~Atwood and A.~Soni,
arXiv:hep-ph/0304085.



\bibitem{sinhas_angle}
D.~London, N.~Sinha and R.~Sinha,
arXiv:hep-ph/0304230;
Phys.\ Rev.\ Lett.\  {\bf 85}, 1807 (2000).

\bibitem{desh}
N. Deshpande and A. Soni, Snowmass Proceedings 1986, p.58
(Eds. R. Donaldson and J. Marx).

\bibitem{dunietz_selftag}
I.~Dunietz,
Phys.\ Lett.\ B {\bf 270}, 75 (1991).





\bibitem{branco} G. Branco, L. Lavoura, J. Silva in
{\it CP Violation}, Oxford Univ. Press (1999), esp. p.445.



\bibitem{kayser}
B.~Kayser and D.~London,
Phys.\ Rev.\ D {\bf 61}, 116013 (2000).





\bibitem{sanda} A.I. Sanda, hep-ph/0108031.  





\bibitem{deltapaper}
D.~Atwood and A.~Soni,
arXiv:hep-ph/0206045.

\bibitem{robf}
R. Fleischer, Nucl. Phys. B659 (2003), 321; Phys. Lett. B 562,
234(2003).  




\bibitem{bkdpi} 
R.~Aleksan, T.~C.~Petersen and A.~Soffer, Phys.\ Rev.\ D {\bf 67}, 096002
(2003).

\bibitem{scs}
Y.~Grossman, Z.~Ligeti and A.~Soffer, hep-ph/0210433;






\bibitem{pdb}
K.~Hagiwara {\it et al.}  [Particle Data Group Collaboration],
Phys.\ Rev.\ D {\bf 66}, 010001 (2002).

\bibitem{prell}
Soren Prell, private communication.

\bibitem{sofferun}
A.~Soffer,
arXiv:hep-ex/9801018.

\bibitem{ss1} J.~P.~Silva and A.~Soffer, Phys.\ Rev.\ D {\bf 61},
112001 (2000).


\bibitem{ggr}
M.~Gronau, Y.~Grossman and J.~L.~Rosner,
Phys.\ Lett.\ B {\bf 508}, 37 (2001).

\bibitem{scs2}
A.~Giri, Y.~Grossman, A.~Soffer and J.~Zupan,
arXiv:hep-ph/0303187.



\bibitem{ckm_fit} 
See, e.g. Atwood and Soni, Phys. Lett. B508, 17 (2001)  






\bibitem{gronau_d}
M.~Gronau,
Phys.\ Rev.\ Lett.\  {\bf 83}, 4005 (1999).




%
%


 
\bibitem{pristine} D.~Atwood and A.~Soni, hep-ph/0212071

\bibitem{my_izu} See the talk by Masa Yamamuchi (BELLE) at
the 5th Workshop on Higher Luminosity B-Factory (Sept 24-26,2003),
Izu, Japan. 

\bibitem{izu} See the Proceedings of 5th Workshop on
Higher Luminosity B-Factory (Sept 24-26,2003), Izu, Japan:
$http:// belle.kek.jp/ superb/ workshop/2003/ HL05/ program.htm$ 

\bibitem{slac} See the Proceedings of the
Second Workshop on the Discovery Potential of an Asymmetric
B-Factory at $10^{36}$ Luminosity (Oct 22-24,2003), SLAC:
http:// www.slac.stanford.edu/ BFROOT/www/ Organization/ 1036\_Study\_Group/ 0310Workshop/


%
\bibitem{E687}
P.~L.~Frabetti {\it et al.}  [E687 Collaboration],
Phys.\ Lett.\ B {\bf 331}, 217 (1994).

\bibitem{cleo_dalitz}
H. Muramatsu et al. (CLEO),
Phys.\ Rev.\ Lett. {\bf 89},251802,2002; Erratum-ibid 90,059901,2003. 

\bibitem{belle_3}
K. Abe et al. (BELLE), hep-ex/0308043.  

\bibitem{mix_formal}
See, e.g. Z. Xing, Phys. Rev. {\bf D55}, 196(1997); T. Liu
{\it Tau Charm Factory Workshop, Argonne IL (1995)}.




\end{thebibliography}
\end{document}